\documentclass[onecolumn,showpacs,preprintnumbers,amsmath,amssymb]{revtex4}

 \usepackage{graphicx}
\usepackage{dcolumn}% Align table columns on decimal point
\usepackage{bm}% bold math

\newcommand{\be}{\begin{eqnarray}}
\newcommand{\ee}{\end{eqnarray}}
\newcommand{\non}{\nonumber\\}

\begin{document}

\title{$S$--wave pion nucleon scattering lengths from $\pi N$, pionic hydrogen and deuteron data}
      
 \author{M. D\"oring}
  \email{michael.doering@ific.uv.es}
 \author{E. Oset} 
  \email{oset@ific.uv.es}
 \author{M.J. Vicente Vacas}
  \email{manuel.j.vicente@uv.es}

\affiliation{
    Departemento de F\'{\i}sica Te\'orica and IFIC, Centro Mixto \\
Universidad de Valencia--CSIC, 46100 Burjassot (Valencia), Spain}

\begin{abstract}
The isoscalar and isovector scattering lengths $(b_0,b_1)$ are determined using a unitarized coupled channel approach based on chiral Lagrangians. Using experimental values of pionic hydrogen and deuterium as well as low energy $\pi N$ scattering data, the free parameters of the model are calculated. Isospin violation is incorporated to a certain extent by working with physical particle masses. For the deuterium scattering length $a_{\pi^- d}$ new significant corrections concerning real and imaginary parts are evaluated, putting new constraints from $\pi^- d$ scattering on the values of $(b_0,b_1)$. In particular, dispersion corrections, the influence of the $\Delta(1232)$ resonance, crossed terms and multiple scattering in a Faddeev approach are considered. 
\end{abstract}


\maketitle

\section{Introduction}
The precise values of the isoscalar and isovector $\pi N$ scattering lengths are one of the important issues in hadronic physics. Together with low energy $\pi N$ scattering data they determine parameters of the chiral Lagrangian which allows to make predictions even below $\pi N$ threshold using chiral perturbation theory. The experimental data from where $(b_0,b_1)$ are usually extracted are the shift and width of pionic hydrogen and deuterium atoms. From the recent measurements at PSI one deduces the elastic $\pi^- p$ plus the $\pi^- p \to \pi^0 n$ transition scattering lengths \cite{Schroder:uq,Schroder:rc,Sigg:qd,Sigg:1995wb}. Using in addition the measured $a_{\pi^- d}$ amplitude \cite{Hauser:yd,Schroder:rc} in order to determine the isospin even and odd combinations
\be
a_+&=&\frac{1}{2}\;\left(a_{\pi^-p}+a_{\pi^-n}\right)\non
a_-&=&\frac{1}{2}\;\left(a_{\pi^-p}-a_{\pi^-n}\right),
\label{first_eqn}
\ee
or, correspondingly, the isoscalar and isovector scattering lengths $b_0=a_+$, $b_1=-a_-$, requires, however, a non--trivial work on the $\pi^- d$ system. This is because the impulse approximation (IA) vanishes in the limit $b_0=0$ and the extraction of $a_{\pi^-n}$ from $a_{\pi^- d}$ calls for a multiple scattering treatment with the double scattering as the leading contribution. Also, higher order corrections as absorption and dispersion have an important effect, as has been extensively discussed for instance in ref. \cite{Ericson:2000md}. It is in particular the $\pi^- d$ scattering length that narrows down the value of $(b_0,b_1)$. Although the error of $a_{\pi^- d}$ is dominated by large theoretical uncertainties, the corrections on $a_{\pi^- d}$ directly affect the values of $(b_0,b_1)$.  

The determination of the pion deuteron scattering length from the elementary ones is one of the problems which has attracted much attention \cite{Weinberg:1966kf,Afnan:1974ye,Mizutani:xw,Fayard:1980xi,Thomas:1979xu} in the past, but has also stimulated more recent studies \cite{Ericson:2000md,Borasoy:gf,Baru:2003yb,Bahaoui:2003xb,Baru:xf,Tarasov:yi}. In ref. \cite{Mizutani:xw}, $a_{\pi^- d}$ is calculated in a Faddeev approach incorporating several processes in the multiple scattering series as nucleon--nucleon correlations, absorption and the corresponding dispersion. It delivers, together with \cite{Afnan:1974ye}, a very complete description of multiple scattering in the deuteron. Here, instead, we use the fixed center approximation (FCA) to the Faddeev equations. Other contributions, like absorption and dispersion correction, are evaluated separately, and fully dynamically, in a Feynman diagrammatic approach. This is feasible because the multiple scattering series is rapidly converging since the scattering lengths are small compared to the deuteron radius. 

In ref. \cite{Beane:2002wk} the values of $(b_0,b_1)$ have been calculated from the pion deuteron scattering length up to NNLO in chiral perturbation theory including the ${\cal L}^{(2)}$ and ${\cal L}^{(3)}$ $\pi N$ Lagrangians. Using realistic deuteron wave functions, and other modifications, the authors reobtain the double and triple scattering formulas in the isospin limit.

Typical results for the isoscalar and isovector scattering lengths, obtained recently in ref. \cite{Ericson:2000md} and ref. \cite{Beane:2002wk}, are:
\be
\left(b_0,b_1\right)=&\left(-12\pm 2 \;\mbox{stat.}\; \pm 8 \;\mbox{syst.}, -895\pm 3 \;\mbox{stat.}\; \pm 13\; \mbox{syst.}\;\right)&\cdot 10^{-4}\;m_{\pi^-}^{-1} \quad\mbox{\cite{Ericson:2000md}}\non
\left(b_0,b_1\right)=&\left(-34\pm 7,-918\pm 13\right)&\cdot 10^{-4}\;m_{\pi^-}^{-1} \quad\mbox{\cite{Beane:2002wk}}.
\label{other_bs}
\ee
Here, and in the subsequent results, the unit of inverse pion mass refers to the charged pion.
The results of eq. (\ref{other_bs}) are not in agreement with each other. The problem with $a_+$ is that it becomes a small quantity from a cancellation of terms of the order of $a_-$, hence $a_+$ is difficult to determine, and the discrepancy between the two results in eq. (\ref{other_bs}) indicates that the uncertainties in $a_+$ are larger than shown in eq. (\ref{other_bs}). Actually, in ref. \cite{Beane:2002wk} larger uncertainties are advocated from isospin violation, since the analysis is made by assuming isospin symmetry. The present study is formulated in the particle base and thus, isospin breaking effects from different physical masses are incorporated.
This provides a part of the isospin violation \cite{Fettes:1998ud,Fettes:1998wf,Gibbs:1995dm,Gibbs:1997jv,Matsinos:1997pb,Piekarewicz:1995tx} that has already been observed in a similar context in $\overline{K}N$ scattering \cite{Oset:1997it}. 

Our first purpose in the present study is to carry out further calculations in the problem of $\pi^- d$ scattering at threshold incorporating novel terms. We start with the absorption of the $\pi^-$ in the deuteron and the dispersion tied to it. The latter contributes to the real part of $a_{\pi^- d}$, and in the literature a quite large correction originates from this source. Since high precision deuteron wave functions are at hand nowadays, and the analysis is carried out fully dynamically, a revision of the results from \cite{Afnan:1974ye,Mizutani:xw,Fayard:1980xi} is appropriate. After calculating the effects of the $\Delta(1232)$ excitation in the dispersion, other contributions as crossed terms are considered. Together with corrections of different nature from the literature, a final correction to $a_{\pi^- d}$ is given. This enables us to parametrize the pion deuteron scattering length in terms of the elementary $s$--wave $\pi N$ scattering lengths $a_{\pi N}$ via the use of the Faddeev equations. We also test the model dependence of the results on the deuteron wave functions. 

The second purpose is then the application of the unitary coupled channel model from ref. \cite{Inoue:2001ip} to $\pi N$ scattering at low energies. The model provides the $\pi N$ scattering lengths for the Faddeev equations for $a_{\pi^- d}$ and the $\pi N$ scattering amplitudes at low energies and threshold. 
First, it is tested if the model can explain threshold data and low energy $\pi N$ scattering consistently. Then, a precise parametrization of the $\pi N$ amplitude at low energies, including threshold, is achieved.

This can be used to extrapolate to the negative energies felt by pionic atoms, the study of which has been  one of the stimulating factors in performing the present work.

\section{Summary of the model for $\pi N$ interaction}
\label{sec_unitary_model}
We follow here the approach of ref. \cite{Inoue:2001ip} where the $N/D$ method adapted to the chiral context of \cite{Oller:1998zr} is applied. Developed for the case of meson meson interactions, the method of \cite{Oller:1998zr} was extended to the meson baryon interaction in \cite{Meissner:1999vr,Oller:2000fj}, and ref. \cite{Inoue:2001ip} follows closely the formalism of these latter works. 
In the CM energy region of interest, from threshold up to around 1250 MeV, pions and nucleons play the predominant role compared to the 
influence of the heavier members of the meson and baryon octet. We have carried out the $SU(3)$ study as in ref. \cite{Inoue:2001ip}, which involves the $K\Sigma$, $K\Lambda$ and $\eta n$ channels in addition to the $\pi N$ ones. At low energies, these channels are far off shell and we have seen that the fit to the data improves only slightly at the cost of three new additional subtraction constants. 
Therefore, we restrict the coupled channel formalism to $\pi^-p$, $\pi^0 n$ in the charge zero sector, and $\pi^+ p$ in the double charge sector. 
The scattering amplitudes are described by the Bethe--Salpeter equation
\be
T\left(\sqrt{s}\right)^{-1}=V^{-1}\;\left(\sqrt{s}\right)-G\left(\sqrt{s}\right)
\label{bs}
\ee
where the kernel $V$ is obtained from the lowest order meson baryon Lagrangian \cite{Meissner:1993ah,Bernard:1995dp,Ecker:1994gg}
\be
V_{ij}\left(\sqrt{s}\right)=-C_{ij}\;\frac{1}{4f_\pi^2}\;\left(2\sqrt{s}-M_i-M_j\right)\;\sqrt{\frac{M_i+E_i\left(\sqrt{s}\right)}{2M_i}}\;\sqrt{\frac{M_j+E_j\left(\sqrt{s}\right)}{2M_j}}
\label{kernel}
\ee
where $C_{ij}$ are the $SU(3)$ coefficients evaluated in ref. \cite{Inoue:2001ip},  
and $G$ is the loop function of the pion nucleon propagator, which in dimensional regularization reads:
\be
G_i\left(\sqrt{s}\right)&=&\frac{2M_i}{(4\pi)^2}\;\Big[\;\alpha(\mu)+\log\frac{m_i^2}{\mu^2}+\frac{M_i^2-m_i^2+s}{2s}\;\log\frac{M_i^2}{m_i^2}\non
&+&\frac{Q_i\left(\sqrt{s}\right)}{\sqrt{s}}\Big[ \log\left(s-\left(M_i^2-m_i^2\right)+2\sqrt{s}\;Q_i\left(\sqrt{s}\right)\right)+ \log\left(s+\left(M_i^2-m_i^2\right)+2\sqrt{s} \;Q_i\left(\sqrt{s}\right)\right)\non
&&- \log\left(-s+\left(M_i^2-m_i^2\right)+2\sqrt{s}\;Q_i\left(\sqrt{s}\right)\right)- \log\left(-s-\left(M_i^2-m_i^2\right)+2\sqrt{s} \;Q_i\left(\sqrt{s}\right)\right)\Big]\Big]\non
\ee 
where $Q_i\left(\sqrt{s}\right)$ is the on shell center of mass momentum of the $i$--th pion nucleon system and $M_i (m_i)$ are the nucleon (pion) masses. The parameter $\mu$ sets the scale of regularization ($\mu=1200$ MeV) and the subtraction parameter $\alpha$ is fitted to the data.
In order to ensure isospin conservation in the case of equal masses, the subtraction constants $\alpha_{\pi^- p}$, $\alpha_{\pi^0 n}$, and $\alpha_{\pi^+ p}$ are taken to be equal for states of the same isospin multiplet. Isospin breaking effects from other sources than mass differences are discussed later (by allowing the $\alpha_{\pi N}$ to be different).

The work of \cite{Inoue:2001ip} concentrated mostly in the region around the $N^\star(1535)$ resonance, which is dynamically generated in the scheme. The data around this region were well reproduced, although the description of the $I=3/2$ sector required the introduction of the extra $\pi\pi N$ channel. The low energy data was somewhat overestimated in \cite{Inoue:2001ip} although qualitatively reproduced. Here, however, our interest is to concentrate around threshold in order to obtain an accurate as possible description of the data in this region and determine, together with the pionic atom data on hydrogen and deuterium, the values of the isoscalar and isovector scattering lengths, with a realistic estimate of the error. 

\subsubsection{Isoscalar Piece}
The chiral Lagrangian at lowest order that we use contributes only to the isovector $\pi N$ amplitude at tree level, but isoscalar contributions are generated from rescattering. Additional isoscalar terms emerge in the expansion in momenta of the chiral Lagrangian \cite{Fettes:1998ud,Meissner:1999vr,Kaiser:1996js}, and we take the relevant terms from ${\cal L}_{\pi N}^{(2)}$ into account following ref. \cite{Fettes:2000bb}. In particular, there is a term independent of $q^0$ with $q$ the pion momentum, and one quadratic in $q^0$, which enter into the potential $V$ from eq. (\ref{kernel}) as
\be
V_{ij}\to V_{ij}+\delta_{ij} \left(\frac{4 c_1-2c_3}{f_\pi^2} \; m_\pi^2-2c_2\;\frac{(q^0)^2}{f_\pi^2}\right)\frac{M_i+E_i\left(\sqrt{s}\right)}{2M_i}.
\label{iso_lagrangian}
\ee
The on shell value of the $c_3$ term ($c_3 q^2$ in ref. \cite{Fettes:2000bb}) has been taken, consistently with the approach of refs. \cite{Inoue:2001ip,Oller:2000fj} which uses the on shell values for the vertices in the scattering equations.
The $c_i$ --combinations, in the notation of ref. \cite{Fettes:2000bb}, are fitted to the experiment. For a construction of a $\pi N$ potential up to higher energies, one has to regulate the quadratic term with $c_2$, which we do by multiplying it by a damping factor,
\be
e^{-\beta^2[(q^0)^2-m_\pi^2]}.
\ee
Both cases, with the damping factor, and without are studied. 

\subsubsection{The $\pi N\to \pi\pi N$ channel}

The $\pi\pi N$ channel opens up at CM energies around $1215$ MeV. In the fits, we include energies higher than that, and therefore the 2--loop diagrams from this source should be taken into account as described in ref. \cite{Inoue:2001ip}. In ref. \cite{Inoue:2001ip}, various functional forms for the real part of the $\pi\pi N$ propagator have been tested, and setting it identically to zero resulted in good data agreement. Here we approximate it as a function constant in energy $\sqrt{s}$ and parametrize it in terms of the quantity $\gamma$. The imaginary part of the $\pi\pi N$ propagator has been calculated explicitly in ref. \cite{Inoue:2001ip}. At the energies of interest, it is small, and becomes only important at higher energies. 

The energy dependence of the $\pi N\to \pi\pi N$ vertices has also been determined in ref. \cite{Inoue:2001ip} from $\pi N\to \pi\pi N$ data. Fig. 12 of that reference shows that at low energies they can be well represented by constants, namely $a_{11}=2.6\;m_\pi^{-3}$ and $a_{31}=5.0\;m_\pi^{-3}$, which are the values we use. In the present approach, the $\pi\pi N$ propagator with its two adjacent $\pi\pi N$ vertices provide $\pi N\to \pi N$ amplitudes which are added directly to the kernel of the Bethe--Salpeter equation (\ref{bs}). With the notation of ref. \cite{Inoue:2001ip}, we obtain for the $\pi N \to \pi N$ channels:
\be
&\pi^-p\to\pi^-p:\;\delta V =& \left[\left(\frac{\sqrt{2}}{3}\;a_{11}+\frac{\sqrt{2}}{6}\;a_{31}\right)^2+ \left(\frac{1}{3}\;a_{11}-\frac{1}{3}\;a_{31}\right)^2\right]\;\gamma\non
&\pi^-p\to\pi^0 n:\;\delta V =& \Bigg[\left(\frac{\sqrt{2}}{3}\;a_{11}+\frac{\sqrt{2}}{6}\;a_{31}\right) \left(-\frac{1}{3}\;a_{11}+\frac{1}{3}\;a_{31}\right)\non
&&+\left(\frac{1}{3}\;a_{11}-\frac{1}{3}\;a_{31}\right) \left(-\frac{\sqrt{2}}{6}\;a_{11}-\frac{\sqrt{2}}{3}\;a_{31}\right)
\Bigg]\;\gamma\non
&\pi^0 n\to\pi^0 n:\;\delta V =& 
\left[\left(\frac{\sqrt{2}}{6}\;a_{11}+\frac{\sqrt{2}}{3}\;a_{31}\right)^2+
\left(\frac{1}{3}\;a_{11}-\frac{1}{3}\;a_{31}\right)^2\right]\;\gamma\non
&\pi^- n\to\pi^- n:\;\delta V =& 
\left(-\sqrt{\frac{1}{2}}\;a_{31}\right)^2\gamma
\label{pipinchannels}
\ee 
where all possible $\pi\pi N$ intermediate states in the loop are considered.
The $\delta V$ of eq. (\ref{pipinchannels}) are then added to the kernel $V_{ij}$ together with the isoscalar piece. In section \ref{sec_results} the constraints of $\gamma$ are discussed: The contribution from the $\pi\pi N$ channels should not exceed a small percentage of the corresponding $V_{ij}$ at $\pi N$ threshold.

\subsubsection{Further refinements of the coupled channel approach}
Following the outline of ref. \cite{Inoue:2001ip} we take into account the Vector Meson Dominance (VMD) hypothesis and let the $\rho$ meson mediate the meson baryon interaction in the $t$ channel. This is justified by the identical coupling structure of the $\rho NN$ coupling within VMD and the kernel $V$ from eq. (\ref{kernel}), thus revealing the lowest order chiral Lagrangian as an effective manifestation of VMD. The $\rho$ meson exchange is incorporated in the formalism via a modification of the coefficients $C_{ij}$ in (\ref{kernel}) --- for details see \cite{Inoue:2001ip}. The explicit consideration of the $\rho$ exchange helps to obtain a better energy dependence, reducing the strength of the amplitudes as the energy increases.

One of the conclusions in \cite{Beane:2002wk} was that the uncertainties of the $(b_0,b_1)$ values should be bigger than quoted in the paper due to the neglect of isospin violation in the analysis. In the present work we introduce a certain amount of isospin violation by working in coupled channels keeping the exact masses of the particles. Although this is not the only origin of isospin violation \cite{Fettes:1998ud,Gibbs:1995dm,Gibbs:1997jv,Matsinos:1997pb,Piekarewicz:1995tx} it gives us an idea of the size of uncertainties from this source. Since there are threshold effects in the amplitudes, and the thresholds are different in different $\pi N$ channels, this leads to non negligible isospin breaking effects as was shown in the case of $\overline{K}N$ interaction in \cite{Oset:1997it}.

\vspace*{0.2cm}

Thus, the parameters for the fit of the $s$--wave amplitude in $\pi N$ scattering are the subtraction constant $\alpha$ from the $\pi N$ loop, two parameters  from the isoscalar terms of the ${\cal L}_{\pi N}^{(2)}$ chiral Lagrangian, and $\gamma$ from the $\pi N\to\pi\pi N$ loop. For the parametrization of the $\pi  N$ potential up to higher energies, a damping factor, introducing another parameter for the quadratic isoscalar term, is studied. In order to account for isospin breaking from other sources than mass splitting, different $\alpha_i$ for the three $\pi N$ channels will be investigated.

\section{Pion deuteron scattering}
The traditional approach to $\pi^- d$ scattering is the use of Faddeev equations \cite{Afnan:1974ye,Mizutani:xw,Fayard:1980xi}, although the fast convergence of the multiple scattering series makes the use of the first few terms accurate enough. On top of this there are other contributions coming from pion absorption, and the dispersion contribution tied to it, crossed terms and the $\Delta(1232)$ resonance, plus extra corrections which are discussed in detail in ref. \cite{Ericson:2000md}.  
\subsection{Faddeev approach}
\label{sec_faddeev}
We follow here the fixed centre approximation (FCA) to the Faddeev equations which was found to be very accurate in the study of $K^- d$ scattering \cite{Kamalov:2000iy} by comparing it to a full Faddeev calculation \cite{Deloff:gc,Deloff:2001zp}. See also the recent discussion endorsing the validity of the static approximation in ref. \cite{Baru:2004kw}. The FCA accounts for the multiple scattering of the pions with the nucleons assuming these to be distributed in space according to their wave function in the deuteron. The Faddeev equations in the FCA are given in terms of the Faddeev partitions
\be
T_{\pi^- d}=T_p+T_n
\ee
where $T_p$ and $T_n$ describe the interaction of the $\pi^-$ with the deuteron starting with a collision on a proton and a neutron respectively. The partitions at threshold satisfy 
\be
T_p&=&t_p+t_pGT_n+t^xGT^x\non
T_n&=&t_n+t_nGT_p\non
T^x&=&t^x+t_n^0GT^x+t^xGT_n
\label{faddeeveqn}
\ee
Here, $G$ is the pion propagator and $t_p$, $t_n$, $t_n^0$, $t^x$ the elementary $s$--wave scattering $T$--matrices of $\pi^-$ on proton and neutron, $\pi^0$ on the neutron, and the charge exchange $\pi^- p\leftrightarrow\pi^0 n$, in this order. While the full Faddeev approach involves integrations over the pion momentum, the FCA factorizes the pion propagator to $G\sim 1/r$ and eqns. (\ref {faddeeveqn}) become a coupled set of algebraic equations. 
These equations are at the level of operators. At any place where charge is transferred from one
nucleon to the other, the sign has to
be changed due to the exchanged final state (Deuteron $\sim \frac{1}{\sqrt{2}}(|pn\rangle-|np\rangle)$). Following ref. \cite{Kamalov:2000iy} we find for the $\pi^- d$ amplitude density
\be
\hat{A}_{\pi^-d}(r)=\frac{\tilde{a}_p+\tilde{a}_n+\left(2\tilde{a}_p\tilde{a}_n-b_x^2\right)/r-2b_x^2\tilde{a}_n/r^2}{1-\tilde{a}_p\tilde{a}_n/r^2+b_x^2\tilde{a}_n/r^3}, \quad b_x=\tilde{a}_x/\sqrt{1+\tilde{a}_n^0/r}
\label{analytic_kamalov}
\ee
with $\tilde{a}_i$ being related to the scattering lengths $a_i$ and the elementary $t_i$  by
\be
\tilde{a}_i=\left(1+\frac{m_\pi}{m_N}\right)a_i=-\frac{1}{4\pi}t_i. 
\label{various_conversions}
\ee
The masses in eq. (\ref{various_conversions}) have to be understood as the physical ones in each channel.

The final $\pi^- d$ scattering amplitude is then obtained by folding the amplitude density with the deuteron wave function as
\be
a_{\pi^- d}=\frac{M_d}{m_{\pi^-}+M_d}\;\int d{\bf r}\;|\varphi_d({\bf r})|^2\hat{A}_{\pi^- d}(r).
\label{folding_a}
\ee
This is the real part of $a_{\pi^- d}$ that has to be modified by the corrections of the following sections. The latter will also provide the correct imaginary part of the pion--deuteron scattering length.

If we keep up to the $(1/r)^2$ terms in eq. (\ref{analytic_kamalov}) and assume isospin symmetry, the resulting formula coincides with the triple scattering result of \cite{Beane:2002wk} up to ${\cal O}(p^4)$ in their modified power counting. In order to show the convergence of the multiple scattering series in the the $\pi^- d$ collision we show in Table \ref{table_one} the different contributions for two cases: First, for the experimental values from ref. \cite{Schroder:rc} with $(b_0,b_1)=(-0.0001,-0.0885)m_{\pi^-}^{-1}$, and second for $(b_0,b_1)=(-0.0131,-0.0924)m_{\pi^-}^{-1}$ from the phenomenological Lagrangian of ref. \cite{Garcia-Recio:1987ik} in eq. (\ref{pheno_la}).
\linespread{1.3}
\begin{table}
\caption{Contributions to the multiple scattering series for Re $a_{\pi^- d}$.}
\begin{tabular*}{0.7\textwidth}{@{\extracolsep{\fill}}|l||l|l|}
%\begin{tabular}{|l||l|l|}
\hline
& from \cite{Schroder:rc}$\left[m_{\pi}^{-1}\right]$&Phenom. Ham. \cite{Garcia-Recio:1987ik}$\left[m_{\pi}^{-1}\right]$ 
\\ \hline
$(b_0,b_1)$&$(-0.0001,-0.0885)$&$(-0.0131,-0.0924)$
\\ \hline
Impulse Approximation&$-2.14\cdot 10^{-4}$&$-0.02793$
\\
Double Scattering&$-0.02527$&$-0.02725$
\\
Triple Scattering&$0.002697$&$0.003489$
\\
4-- and higher scattering&$1.06\cdot 10^{-4}$&$5.4\cdot 10^{-5}$
\\ \hline
Solution Faddeev&$-0.02268$&$-0.05163$
\\ \hline
\end{tabular*}
\label{table_one}
\end{table}
\linespread{1.0} We can see that in both cases the double scattering is very important and in the case of \cite{Schroder:rc} where $b_0$ is quite small, the double scattering is the leading contribution.

\subsection{Absorption and dispersion terms}
\label{sec_absorption_dispersion}
Pion absorption in deuterium has been studied in \cite{Maxwell:xm} using Feynman diagrammatic techniques. The absorption contribution reflects into the imaginary part of the (elastic) $\pi^- d$ scattering length. Its diagrammatic evaluation leads at the same time to a dispersive real contribution to the $\pi^- d$ scattering length. The evaluation of this latter contribution has been done using Faddeev approaches \cite{Afnan:1974ye,Mizutani:xw,Thomas:1979xu}. We shall evaluate it here including extra terms from the $\Delta$ excitation, going beyond the non--relativistic treatment of the pions in \cite{Afnan:1974ye,Mizutani:xw}, and testing various approximations for the dispersive part. 

\begin{figure}
\begin{picture}(200,220)
\put(-80,10){
\includegraphics[width=12cm]{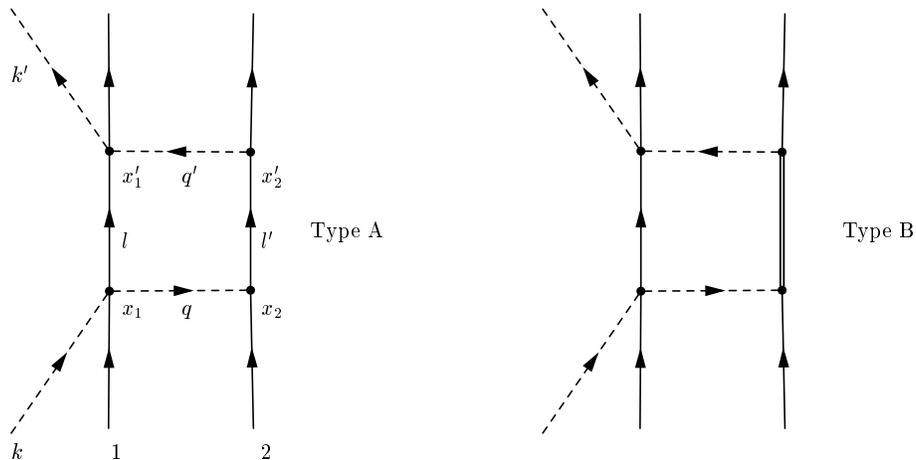}}
\end{picture}
\caption{Absorption plus dispersion terms in $\pi^- d$ scattering.}
\label{absorption_diagram}
\end{figure}

\begin{figure}
\begin{picture}(200,160)
\put(-150,0){
\includegraphics[width=16cm]{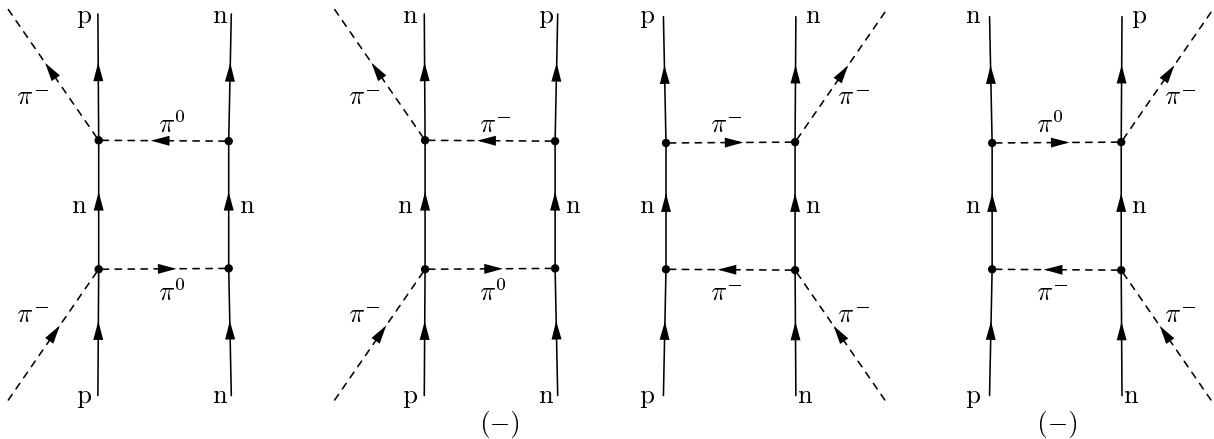}}
\end{picture}
\caption{Charge States in Absorption.}
\label{charge_states_in_absorption}
\end{figure}

\begin{figure}
\begin{picture}(200,160)
\put(-150,0){
\includegraphics[width=16cm]{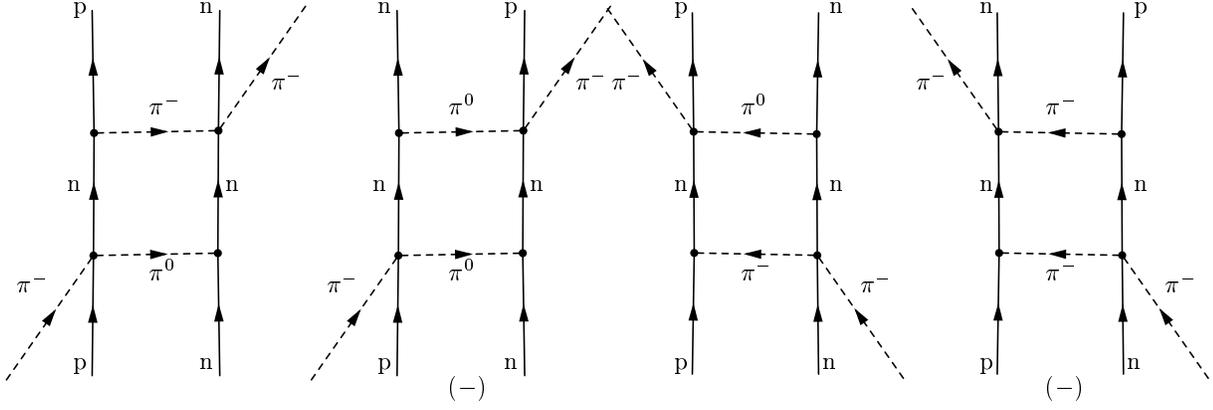}}
\end{picture}
\caption{Additional diagrams in absorption.}
\label{charge_man}
\end{figure}

In order to evaluate the absorption and dispersion terms, a Feynman diagrammatic approach is used which offers flexibility to account for different mechanisms. We shall evaluate the contribution of the diagrams of Fig. \ref{absorption_diagram},  where Type B contributes only to the real part,
including permutations of the scattering vertices on different nucleons and different time orderings as shown in Figs. \ref{charge_states_in_absorption} and \ref{charge_man}.

On the first hand we consider the diagrams of type A and find the possibilities shown in Fig. \ref{charge_states_in_absorption}.
For the purpose of evaluating the absorption and dispersion corrections we shall use the effective Hamiltonian \cite{KoRe1966,Garcia-Recio:1987ik}
\be
H_I=4\pi\left[\frac{\lambda_1}{m_\pi}\;\overline{\Psi}\vec{\phi}\vec{\phi}\Psi+\frac{\lambda_2}{m_\pi^2}\; \overline{\Psi}\vec{\tau}(\vec{\phi}\times\partial^0\vec{\phi})\Psi\right]
\label{pheno_la}
\ee
with $\lambda_1=0.0075$, $\lambda_2=0.053$, which shows the dominance of the isovector part with $\lambda_2$. For the $\pi NN$ vertex the usual Yukawa $\left(f_{\pi NN}/m_\pi\right)\;{\bf \sigma \cdot q}\;\tau^\lambda$  vertex is taken. The value of $\lambda_2$ corresponds very closely to the final isovector term that we find, while the value of $\lambda_1$ is about twice as large. Yet, using the new values that come from our analysis in a first step of a selfconsistent procedure only leads to changes in the final results that are much smaller than the uncertainties found. 

The normalization of our $T$ amplitude is such that the scattering matrix $S$ is given by
\be
S=1-i\;\frac{1}{V^2}\frac{1}{\sqrt{2\omega}}\frac{1}{\sqrt{2\omega'}}\sqrt{\frac{M_d}{E_d}}\sqrt{\frac{M_d}{E'_d}}\;T\;\left(2\pi\right)^4 \delta^4\left(p_{\pi^-}+p_d-p'_{\pi^-}-p'_d\right).
\ee
From the diagram A of Fig. \ref{absorption_diagram} we obtain in the $\pi^- d$ center of mass frame after performing the $q^0$ and $q'^{0}$ integrations
\begin{eqnarray}
T&=&i\int\frac{d^4l}{(2\pi)^4}\int\frac{d^3{\bf q}}{(2\pi)^3}\int\frac{d^3{\bf q'}}{(2\pi)^3}\; F_d({\bf q}+{\bf l})F_d({\bf q'}+{\bf l})\nonumber\\
&&\frac{1}{q^2-m_\pi^2+i\epsilon}\;\frac{1}{q'^2-m_\pi^2+i\epsilon}\;\frac{1}{l^0-\epsilon({\bf l})+i\epsilon}\;\frac{1}{l'^0-\epsilon({\bf l'})+i\epsilon}\nonumber\\
&&\Sigma t_1\;t_2\;t_{1'}\;t_{2'}\;\left({\bf q \cdot q'}\right)
\label{t_absorption}
\end{eqnarray}
with $q=\left(m_\pi-l^0,{\bf q}\right),\;q'=\left(m_\pi-l^0,{\bf q}'\right),{\bf l}'=-{\bf l}$, and $\epsilon({\bf l})$ refers to the nucleon kinetic energy. In eq. (\ref{t_absorption}), $F_d$ is the deuteron wave function in momentum space including $s$-- and $d$--wave (see Appendix), and the amplitude from the sum of diagrams of Fig. \ref{charge_states_in_absorption} is given by
\be
\Sigma t_1\;t_2\;t_{1'}\;t_{2'}=2 \left(4\pi\right)^2\;\frac{1}{m_\pi^2}\left(2\lambda_1+3\lambda_2\right)^2\left(\frac{f_{\pi NN}}{m_\pi}\right)^2 \simeq 40.0\;{\rm fm}^4
\label{absiso}
\ee
where we have made the usual approximation that $q^0$ and $q'^{0}$ in the $\pi N\to \pi N$ amplitude are taken as $m_\pi /2$ which is exact for $\mbox{Im}\; T$.

The ${\bf q\cdot q'}$ term in eq. (\ref{t_absorption}) comes from the $\pi NN$ $p$--wave vertices $\vec{\sigma}{\bf q'}\;\vec{\sigma}{\bf q}={\bf q'\cdot q}+i\left({\bf q}'\times{\bf q}\right)\vec{\sigma}$ after neglecting the crossed product term which does not contribute when using the $s$--wave part of the deuteron wave function. 
 
A different topological structure for the absorption terms is possible and given by the diagrams shown in Fig \ref{charge_man}.
The evaluation of these diagrams involves now the spin of both the nucleons 1 and 2 and one obtains the combination
\be
\sigma_{1i}\sigma_{2j}q_iq'_j
\label{sigma_eins}
\ee
which upon integration over ${\bf q}$, ${\bf q'}$ leads to a structure of the type
\be
\sigma_{1i}\sigma_{2j}l_il_j
\label{sigma_zwei}
\ee
for the $s$--wave part of the wave function.
The extra $l$ integration, involving $l_il_j$ and terms with even powers of ${\bf l}$ allows one to write
\be
\sigma_{1i}\sigma_{2j}l_il_j \longrightarrow\frac{1}{3}\;\sigma_{1i}\sigma_{2j} {\bf l}^2\delta_{ij}
=\frac{1}{3}\;\vec{\sigma}_1\vec{\sigma}_2{\bf l}^2\equiv\frac{1}{3}\;{\bf l}^2
\label{sigma_drei}
\ee
where in the last step we have used that $\vec{\sigma}_1\vec{\sigma}_2=1$ for the deuteron. The final result leads to $1/3$ of the former contribution from the diagrams of Fig. \ref{charge_states_in_absorption} for the imaginary part. 

The integration over the energy variable $l^0$ in eq. (\ref{t_absorption}) has been performed in three different ways. While for all three calculations the imaginary part stays the same as expected, the real part varies significantly, as we will see in the following. The result for the dispersive part for the diagrams of Figs. \ref{charge_states_in_absorption} and \ref{charge_man} depends much on the treatment of the pion poles and the pion propagator. The choice of the wave function will have only moderate influence on the results.

In a first approximation (App1), only the nucleon pole is picked up in the $l^0$--integration of eq. (\ref{t_absorption}). Furthermore, the energy components $q^0$ and $q'^{0} $ of the pion momenta in the propagators are replaced by the on--shell value of $l^0$ which is $m_\pi/2$. This is exact for the imaginary part of the elastic $\pi^- d$ scattering length. The imaginary part is given by cutting the two internal nucleon lines in the diagrams of Fig. \ref{charge_states_in_absorption} and Fig. \ref{charge_man}, and then putting the two nucleons on--shell. The pion mass in this picture is shared between the two nucleons that obtain a kinetic energy of $m_\pi/2$ each after the absorption of the virtual pion on the second nucleon.   

In a second approach (App2), the pion poles of negative energy in the lower $l^0$ half plane are still neglected, but for $q^0$ and $q'^{0} $ we substitute now the residue of the nucleon pole
\be
\frac{1}{\left(\frac{m_\pi}{2}\right)^2-{\bf q}^2-m_\pi^2}\longmapsto\frac{1}{\left[m_\pi-\epsilon({\bf l})\right]^2-{\bf q}^2-m_\pi^2}.
\ee
This leads to new poles in the integration of eq. (\ref{t_absorption}) which correspond to cuts that affect one pion and one nucleon line of the loops in Figs. \ref{charge_states_in_absorption} and \ref{charge_man}. From kinematical reasons the particles cannot go on--shell for these cuts. Indeed, if also the pion poles of negative energy are taken into account (App3), these poles cancel. 

Approach 3 (App3) makes no simplifications in the $l^0$--integration any more, except the substitution of $q^0=q'^{0}=m_\pi/2$ in the elementary scattering length as in eq. (\ref{absiso}).
The 9--dimensional integral of the amplitude (\ref{t_absorption})  for the diagrams in Fig. \ref{charge_states_in_absorption} in the formulation of approach 3 (App3) is:
\be
T&=&\frac{1}{(2\pi)^9}\int d^3{\bf l}\;d^3{\bf q'}\;d^3{\bf q}\; F_d({\bf q+l}) F_d({\bf q'+l})\;{\bf qq'}\;A\;\Sigma t_1\;t_2\;t_{1'}\;t_{2'},\non
A&=&-\frac{1}{2\epsilon(l)-m_\pi-i\epsilon}\;\frac{(2\epsilon(l)-m_\pi)(\epsilon(l)+\omega)(\epsilon(l)-m_\pi+\omega)+(2\epsilon(l) -m_\pi+2\omega)(\omega'^{2}+\omega' (2\epsilon(l)-m_\pi+\omega))}{2\omega\omega'(\omega+\omega')(\epsilon(l)+\omega)(\epsilon(l)+\omega')(\epsilon(l)-m_\pi+\omega) (\epsilon(l)-m_\pi+\omega')}.\non
\label{all_in_one}
\ee
It can be factorized to integrals of lower dimension by writing the sum of pion energies in the denominator of (\ref{all_in_one}) as  
\be
\frac{1}{\omega(q)+\omega(q')}=\int\limits_0^\infty dx\; e^{-\omega(q)x}\;e^{-\omega(q')x},
\ee
thus simplifying the numerical evaluation. The amplitude in eq. (\ref{all_in_one}), divided by 3, provides the imaginary part of the diagrams in Fig. \ref{charge_man}, whereas the real part of the diagrams in Fig. \ref{charge_man} has a different analytical structure. The diagrams of Fig. \ref{charge_man} contribute with 36\%
to the real part with respect to the diagrams of Fig. \ref{charge_states_in_absorption}.

\linespread{1.3}
\begin{table}[h]
\caption{Real and imaginary contributions from absorption to $a_{\pi^- d}$ for three different approaches. All values in $10^{-4}\cdot m_{\pi^-}^{-1}$.}
\begin{tabular*}{0.9\textwidth}{@{\extracolsep{\fill}}|l||l|l|l|}
%\begin{tabular}{|l||l|l|l|}
\hline
&(App1)&(App2)&(App3) 
\\ \hline
Im $a_{\pi^- d}$, $s$--wave&$57.4\pm 5.7$&idem&idem
\\
Im $a_{\pi^- d}$, $d$--wave&$2.21\pm 0.33$&idem&idem
\\
Im $a_{\pi^- d}$ $s+d$--wave&$59.6\pm 5.3$&idem&idem
\\ \hline
Im $a_{\pi^- d}$ experimental&$63\pm 7$&idem&idem
\\ \hline \hline
$\Delta$ Re $a_{\pi^- d}$, $s$--wave&$19.3\pm 8.2$&$13.6\pm 8.9$&$2.4\pm 4.3$
\\ \hline
\end{tabular*}
\label{table_absorption}
\end{table}
\linespread{1.0}
Table \ref{table_absorption} shows the result of all three approaches for the sum of the diagrams from Fig. \ref{charge_states_in_absorption} and Fig. \ref{charge_man}. We have used two refined wave functions, the CD--Bonn potential in the recent version from ref. \cite{Machleidt:2000ge}, and the Paris potential from ref. \cite{Lacombe:eg}. We take the average of the results obtained with either wave function. The difference of the results gives the error in Table \ref{table_absorption}. The statistical error from Monte--Carlo integrations has been kept below $0.1\cdot 10^{-4}\cdot m_{\pi^-}^{-1}$. The dispersive contribution from the $d$--wave has been only calculated for the CD--Bonn potential from ref. \cite{Machleidt:2000ge}, for the amplitude of approach 3, eq. (\ref{all_in_one}). The numerical value is 
\be
\Delta\; \mbox{Re}\; a_{\pi^- d},\;d-\mbox{wave,}\;\mbox{absorption}=0.18\cdot 10^{-4}\cdot m_{\pi^-}^{-1}.
\nonumber
\ee
We also show the influence of the ${\bf q'\times q}$ term, that stems from the $\vec{\sigma}{\bf q'}\;\vec{\sigma}{\bf q}$ structure of the absorption diagrams of Fig. \ref{charge_states_in_absorption}. It had been omitted in eq. (\ref{t_absorption}), since it contributes only in the $d$--wave $\to$ $d$--wave transition. The contribution  from this source has been calculated for the amplitude of approach 3, in eq. (\ref{all_in_one}), for the CD--Bonn potential. We obtain:
\be
\Delta a_{\pi^- d,\;{\bf q'\times q}}=(0.13-i \;0.38)\cdot 10^{-4}\cdot m_{\pi^-}^{-1}.
\nonumber
\ee
We have also tested the relevant contributions of the absorption process with a Hulthen wave function in two different parametrizations taken from \cite{Kharzeev:2002ei}. The results for both parametrizations would lead to large errors of the order of 40\% for the value of Im $a_{\pi^- d}$ for the $s$--wave in Table \ref{table_absorption}, and even larger ones for the imaginary part from the $d$--wave. This is, because the two $p$--wave vertices make the absorption in the diagrams of Figs. \ref{charge_states_in_absorption} and \ref{charge_man} sensitive to the derivative of the used wave function, and the Hulthen wave function is known to be less accurate, as has also been pointed out in ref. \cite{Ericson:2000md}. Therefore we do not use this simplified wave function in this study.

Whereas the imaginary part in Table \ref{table_absorption} remains the same, the dispersive contribution from the $s$--wave decreases when going from approach 1 to 3. The fact that in (App3) it even changes sign for the Paris potential compared to the CD--Bonn potential is due to cancellations between terms, which by themselves are of larger magnitude. 

\begin{figure}
\begin{picture}(200,300)
\put(-100,0){
\includegraphics[width=13cm]{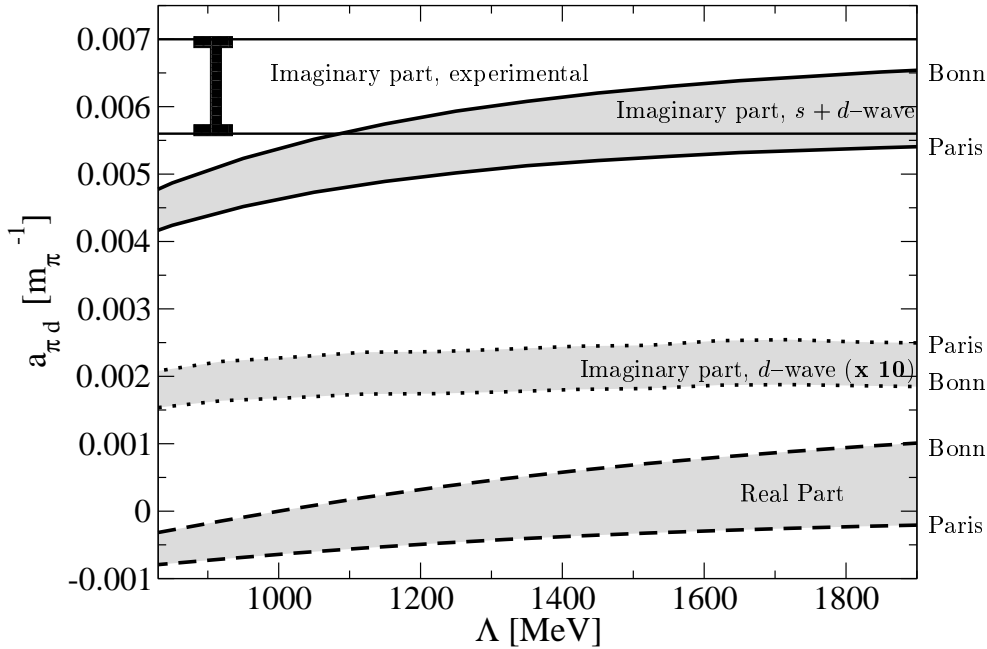}}
\end{picture}
\caption{Real and imaginary contribution to $a_{\pi^- d}$ from absorption as a function of $\Lambda$ from the monopole form factor.}
\label{fig_lambda_dep}
\end{figure}

In all calculations a monopole form factor with cut--off $\Lambda$ has been applied to the $\pi NN$ vertices of the absorption diagrams of Fig. \ref{charge_states_in_absorption} and Fig. \ref{charge_man}. Since $\Lambda$ is the only free parameter involved in the calculation, we plot the dependence on $\Lambda$ of the imaginary part, $s$-- and $d$--wave, and the real part, $s$--wave in Fig. \ref{fig_lambda_dep}. The values correspond to approach 3, eq. (\ref{all_in_one}).

The real part from the $d$--wave is not plotted separately, since it is even one order of magnitude smaller than the imaginary part from the $d$--wave. Also plotted is the experimental value of $\mbox{Im}\;a_{\pi^- d}=0.0063\pm 0.0007 m_{\pi^-}^{-1}$ taken from ref. \cite{Hauser:yd}. The imaginary part from the $d$--state has been amplified ten times in the figure.
 
The results for the imaginary part depend very moderately on $\Lambda$. In Table \ref{table_absorption}, the values correspond to $\Lambda=1.72 \;\mbox{GeV}$ that has been used in the construction of the CD--Bonn potential \cite{Machleidt:2000ge}. We do not observe an amplification of the $d$--wave in absorption, relative to its weight in the deuteron wave function, as claimed in ref. \cite{Mizutani:xw}. From the ${\bf q\cdot q'}$--structure of the absorption amplitude (\ref{all_in_one}) one would expect an amplification of the $d$--wave, which has more weight at higher momenta than the $s$--wave (despite its small contribution to the norm of the deuteron wave function of 4--6\%). However, the correct combination of the angular momentum $l=2$ of the $d$--wave, together with the spin of the nucleons in order to give a total spin of 1, leads to a very effective suppression of this enhancement. 

Also, the effect of rescattering in absorption has been investigated. In order to avoid double counting we consider all rescattering diagrams that have exactly one absorption insertion of the form of the diagrams in Figs. \ref{charge_states_in_absorption} and \ref{charge_man}. In practice, this means a replacement of the two $s$--wave $\pi N$--vertices in the diagrams of Figs. \ref{charge_states_in_absorption} and \ref{charge_man} by a Faddeev--like rescattering series similar to eq. (\ref{faddeeveqn}). By doing so, there are no pion exchanges between the nucleons, which are unconnected to the external pions. Thus, effects, that are already contained in the deuteron wave function, are not double counted.  The explicit evaluation of this class of diagrams results in negligible changes of the values in Table \ref{table_absorption} of the order of 1\% or less.

No interference between $s$ and $d$--wave is observed for absorption and dispersion. In the next section and in the Appendix this issue is discussed further.
 
\subsection{Further corrections to the real part of $a_{\pi^- d}$} 
\label{section_further}
The diagram of Fig. \ref{absorption_diagram} where the nucleon pole is substituted by the $\Delta$ pole (Type B) is evaluated in a similar fashion as Type A. The sum of all possible charge configurations provides now
\be
\Sigma t_1\;t_2\;t_{1'}\;t_{2'}=\frac{32 f_{\pi N\Delta}^2\pi^2\left(4\lambda_1-3\lambda_2\right)^2}{9m_\pi^4}\simeq 10.8 \;{\rm fm}^4
\ee
where $f_{\pi N\Delta}=2.01 f_{\pi NN}$ is the $\pi N\Delta$ $p$--wave coupling. There is no imaginary part from this diagram as the $\Delta$ cannot go on shell. The numerical value can be found as '$\Delta$ excitation' in Table \ref{table_additional}. It is relatively small since the effect of the strong $\pi N\Delta$ coupling is suppressed partly by a cancellation of the isovector part $\sim\lambda_2$ from different charge states for the diagram.

\begin{figure}
\begin{picture}(200,200)
\put(-140,10){
\includegraphics[width=16cm]{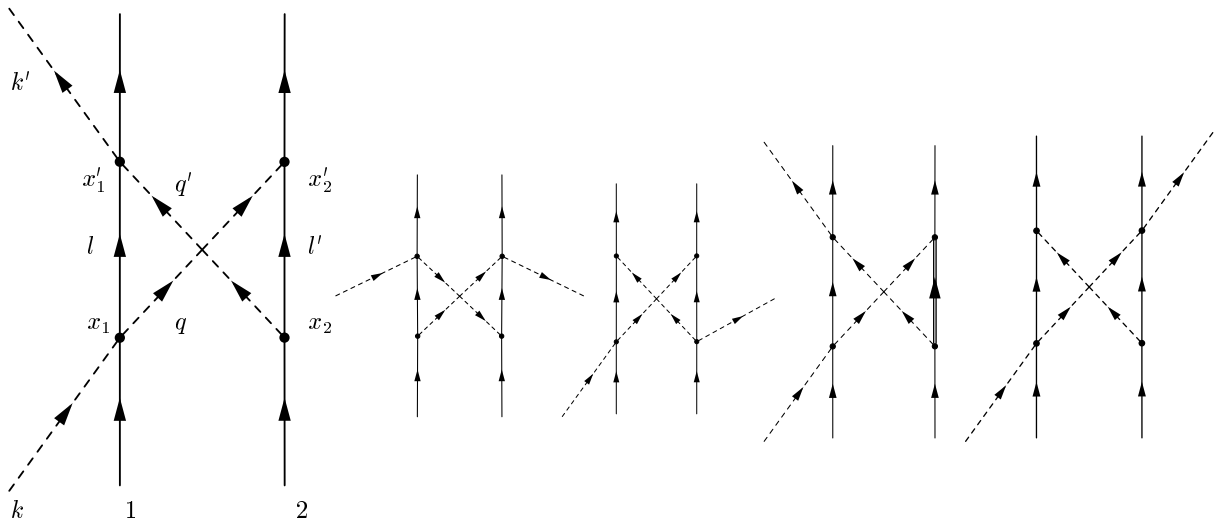}}
\end{picture}
\caption{Crossed Diagrams.}
\label{Crossed_Diagram}
\end{figure}

\linespread{1.3}
\begin{table}
\caption{Values of further contributions to the real part of $a_{\pi^- d}$.}
\begin{tabular*}{0.7\textwidth}{@{\extracolsep{\fill}}|l|l|l|}
%\begin{tabular}{|l|l|l|}
\hline
&{\bf Diagram}&{\bf Value} in $10^{-4}\cdot m_{\pi^-}^{-1}$
\\ \hline
$\Delta$ excitation&Fig. \ref{absorption_diagram}, Type B& $6.4$ 
\\ \hline
Crossed pions&Fig. \ref{Crossed_Diagram}, $1^{\mbox{st}}$&$-1.3\;\pm\;0.1$ (stat.)
\\ \hline
Crossed $\Delta$ excitation& Fig. \ref{Crossed_Diagram}, $4^{\mbox{th}}$&$9.5\;\pm\;1.1$ (stat.)
\\ \hline \hline
Wave function correction (WFC), $s$--wave& Fig. \ref{Crossed_Diagram}, $5^{\mbox{th}}$&$-16.2\;\pm\;0.1$ (stat.)
\\ \hline
WFC, $s$--$d$ interference, $\mu=0$, $\mu'=0$& Fig. \ref{Crossed_Diagram}, $5^{\mbox{th}}$&$14.4\;\pm\;0.1$ (stat.)
\\ \hline
WFC, $s$--$d$ interference, $\mu=0$, $\mu'=\pm 1$& Fig. \ref{Crossed_Diagram}, $5^{\mbox{th}}$&$21.9\;\pm\;0.1$ (stat.)
\\ \hline
\end{tabular*}
\label{table_additional}
\end{table}
\linespread{1.0}
Another source of contribution for the real part of the $\pi^- d$ scattering length is given by the crossed pion diagram displayed in Fig. \ref{Crossed_Diagram}, first diagram.
The $T$ matrix for this process is given by
\be
T&=&\int\frac{d^3{\bf l}}{(2\pi)^3}\frac{d^3{\bf q }}{(2\pi)^3}\frac{d^3{\bf q'}}{(2\pi)^3} \;F_d({\bf q+l})\;F_d({\bf
q'+l})\;\Sigma\left(t_1\;t_2\;t'_1\;t'_2\right)\;\left(\vec{\sigma}\cdot{\bf q}\;\vec{\sigma}\cdot{\bf q'}\right)\non &&
\;\frac{\left(\epsilon+\omega\right)\left(\epsilon'+\omega\right)+\left(\epsilon+\epsilon'+\omega\right)\omega'+\omega'^{2}-m_\pi\left( \epsilon' +\omega+\omega'
\right)}{2\omega\omega'\left(\epsilon'+\omega\right)\left( \epsilon'+\omega'\right)\left(\omega+\omega'\right) \left( m_\pi-\epsilon-\omega\right)
\left(m_\pi-\epsilon-\omega'\right)} 
\label{eval_crossed}
\ee
where the spin--isospin factor
\be
\Sigma\left(t_1\;t_2\;t'_1\;t'_2\right)=8(4\pi)^2 \frac{1}{m_\pi^2}\left(\lambda_1^2-3\;\lambda_1\lambda_2\right)\left( \frac{f_{\pi NN
}}{m_\pi}\right)^2\simeq \;-6.0 \;{\rm fm}^4
\ee
is proportional to the isoscalar scattering length $b_0\sim\lambda_1$, and therefore the contribution of the diagram is very small. In order to show how a non--vanishing $b_0$ influences the crossed diagram, we have evaluated the amplitude in eq. (\ref{eval_crossed}), and the numerical result at the relatively large value of $\lambda_1=0.0075$ leads to the small contribution displayed in Table \ref{table_additional} as 'Crossed pions'. The second and third diagram of Fig.
\ref{Crossed_Diagram} force the nucleons to be far off shell and are equally negligible.

The crossed contribution with the $\Delta$ resonance in the fourth diagram of Fig. \ref{Crossed_Diagram} is similar to
eq. (\ref{eval_crossed}), the only difference being the nucleon kinetic energy $\epsilon'$ which is substituted by $M_\Delta-M_N+{\bf l'}^2/(2M_\Delta)$, and the
spin--isospin factor changing to
\be
\Sigma\left(t_1\;t_2\;t'_1\;t'_2\right)=(4\pi)^2\left[6\lambda_2^2+\frac{16}{3}\;\lambda_1\lambda_2+\frac{32}{9}\;\lambda_1^2\right]\frac{1}{m_\pi^2} \left(\frac{f_{\pi N\Delta}^\star}{m_\pi}\right)^2 \simeq 56\;{\rm fm}^4.
\ee
The diagram provides a larger contribution than the former ones, since it has a large weight in the isovector part $\lambda_2$. This explains why the contribution of this diagram, displayed in Table \ref{table_additional} ('Crossed $\Delta$ excitation'), is relatively large compared to the $\Delta$ excitation in Fig. \ref{absorption_diagram}, Type B, and the crossed pion diagram in Fig. \ref{Crossed_Diagram}, first diagram.

Another type of correction is displayed in the last, fifth diagram of Fig. \ref{Crossed_Diagram}. The difference to the other corrections discussed so far concerns a pion that is not connected to the external pion line. Nevertheless, we are not double counting effects of the deuteron wave function and effects of the diagram. On the contrary, the last diagram of Fig. \ref{Crossed_Diagram} can be understood as a correction of the nucleon--nucleon interaction to double scattering, usually called wave function correction (WFC). The nucleon--nucleon interaction is, of course, modeled by a richer structure in terms of meson exchange as it is considered in the construction of the CD--Bonn potential of ref. \cite{Machleidt:2000ge}. Yet, with the nucleons in the deuteron being relatively far away from each other, one pion exchange should give the right size of this correction in $\pi^- d$ scattering. As one sees in Table \ref{table_additional} ('Wave function correction (WFC), $s$--wave'), the contribution of the $s$--wave of this diagram is less than 1/10 of the one of double scattering, with the same sign (see Table \ref{table_one}). To conclude, the correction induced by the fifth diagram of Fig. \ref{Crossed_Diagram} results in minor changes, that are of the size of 1/3 to 1/2 of triple scattering (see Table \ref{table_one}).
However, we do not include the contribution of this diagram in the determination of the corrections of $a_{\pi^- d}$. This is because it represents part of the non--static effects discussed in the next section.

Besides the contributions of $s$ and $d$--wave to the various corrections discussed so far, the interference between $s$ and $d$--wave should be carefully analyzed. In ref. \cite{Mizutani:xw} sizable contributions from this source were found. The spin structure of the two nucleons together with the angular momentum of the nucleons in the $d$--state prohibits any kind of interference for the diagrams of Fig. \ref{charge_states_in_absorption} and \ref{charge_man}, as an explicit calculation shows. Some explicit formulas for the angular structure of the interference can be found in the Appendix.

The situation is different for the fifth diagram in Fig. \ref{Crossed_Diagram}. There, the pion acts similarly as in the one--pion--exchange, that mixes the small amount of $d$--wave to the $s$--wave of the deuteron wave function. From the $p$--wave character of the $\pi NN$ coupling we expect even an amplification of the interference of $s$ and $d$--wave. This is indeed the case, as the numerical results in Table \ref{table_additional} show. There, we distinguish between interference that leaves the third component of the angular momentum, $\mu$ and $\mu'$ (for incoming and outgoing state), unchanged ('WFC, $s$--$d$ interference, $\mu=0$, $\mu'=0$'), and the interference that involves different values of $\mu$ and $\mu'$  ('WFC, $s$--$d$ interference, $\mu=0$, $\mu'=\pm 1$'). The latter implies a spin flip of one or both nucleons. There is some cancellation between the $s$--wave  and the $s$--wave $d$--wave interference and the net effect is similar to what has been customarily taken as non--static effects in other works, as we mention in the next section.

In addition to the diagrams considered so far we could add others of the type 

\be 
\includegraphics[width=1.5cm]{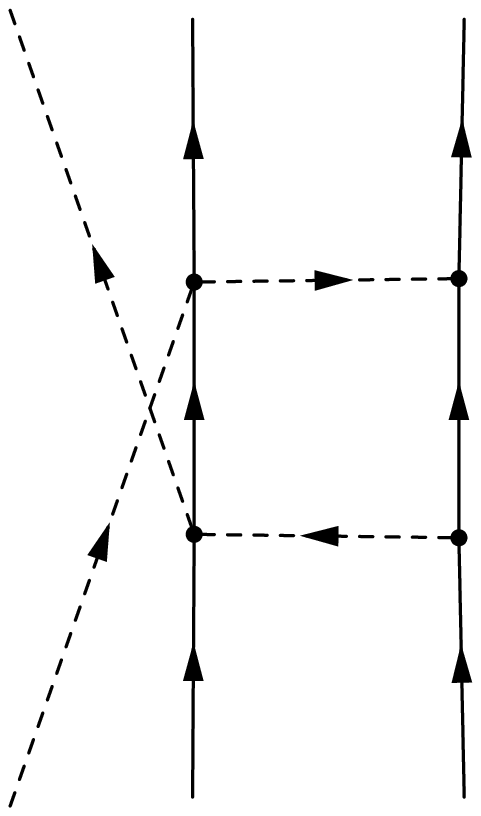} 
\label{crossed_external}
\ee

which also contribute only to the real part. The approximate contribution $\lambda_2\left(m_\pi+m_\pi/2\right)$ in the $\pi N\to \pi N$ $s$--wave vertices becomes now $\lambda_2\left(-m_\pi+m_\pi/2\right)$ and this gives a factor 9 reduction with respect to the other terms plus an extra reduction from the intermediate nucleon propagator and we disregard them.

\subsection{Other Corrections}
\label{corrections}
Besides the dispersive contribution to the real part, the various crossed terms, and the $\Delta$ contribution, all of them discussed in the last section, corrections of a different nature occur for the real part of the $\pi^- d$ scattering length. They have been summarized in ref.
\cite{Ericson:2000md}, and we follow here this work in order to discuss the incorporation of these effects in the present study. 

\vspace*{0.3cm}

{\it Fermi motion / Boost Correction}

\vspace*{0.1cm}

The single scattering term for the $\pi N$ $p$--wave interaction gives a contribution when the finite momentum of the nucleons in the deuteron is taken into account. A value of $61(7)\cdot 10^{-4}m_\pi^{-1}$ arises from the study of ref. \cite{Ericson:2000md} tied to the $c_0$ coefficient of the isoscalar $p$--wave $\pi N$ amplitude. For the $s$--wave such contributions cancel with other binding terms according to ref. \cite{Faldt:1974sm}. In a different analysis in ref. \cite{Beane:2002wk} a finite correction arises from the $s$--wave interaction which is tied to the $c_2$ coefficient in the chiral expansion. The full $c_2$ term in ref. \cite{Bernard:1995dp} provides a momentum dependence which accounts for $p$--wave plus also effective range corrections of the $s$--wave. A large fraction of the $c_2$ coefficient in \cite{Bernard:1995dp} is accounted for by the explicit consideration of the $\Delta$ in \cite{Bernard:1996gq,Fettes:2000bb}. 
The contribution to the $\pi^- d$ scattering length in ref. \cite{Beane:2002wk} depends much on the prescription taken for the expansion and in the NNLO expansion it gives one order of magnitude bigger contribution than the NNLO$^\star$ expansion. The value of the NNLO$^\star$ expansion is considered more realistic in ref. \cite{Beane:2002wk}. We adopt here the scheme followed so far, looking into different mechanisms and using the Bonn and Paris wave functions to have an idea of uncertainties. For this purpose we evaluate the contribution due to Fermi motion in $s$ and $p$--waves.

By using the formalism of ref. \cite{Ericson:gk} we have up to $p$--waves for the $\pi N$ scattering amplitude
\be
{\mathcal F}=b_0+b_1({\tilde{\bf t}}\cdot{\vec\tau})+[c_0+c_1({\tilde{\bf t}}\cdot{\vec\tau})]\;{\bf q'\cdot q}.
\label{co_definition}
\ee
In addition a range dependence of the $s$--wave amplitude is used in \cite{Ericson:2000md} where
\be
a_{\pi^- p}(\omega)&=&a^++a^-+\left(b^++b^-\right){\bf q}^2\non
a_{\pi^- n}(\omega)&=&a^+-a^-+\left(b^+-b^-\right){\bf q}^2
\label{range_expansion}
\ee
By performing a boost to the $\pi N$ CM frame where eqns. (\ref{co_definition},\ref{range_expansion}) hold, we obtain the impulse approximation for the $p$--wave contribution, including the correction for the range of the $s$--wave part as:
\be
a_{\pi^- d}=2\left(c_0+b^+\right)\frac{1+m_\pi/m_N}{1+m_\pi/m_d}\left(\frac{m_\pi}{m_\pi+m_N}\right)^2
\left\langle\left[1+\left(\frac{{\bf p}}{m_\pi+m_N}\right)^2\right]{\bf p}^2\right\rangle
\label{fermi_IA} 
\ee
which coincides with ref. \cite{Ericson:2000md} up to small corrections of ${\cal O}(p^4)$ and the introduction of the $s$--wave range parameter correction which amounts to a 25 \% 
decrease of the term \footnote{T.E.O. Ericson, private communication.}. We obtain from eq. (\ref{fermi_IA}) a contribution of $57\pm 9\cdot 10^{-4}m_\pi^{-1}$ with 9 \%
of this value coming from the ${\bf p}^4$ term.

Next, we also take into account corrections due to double scattering with one $s$--wave and another $p$--wave, or two $p$--waves. This leads to small corrections but with large uncertainties which are genuine and should be taken into account as we show below. Another reason to explicitly evaluate these corrections is that they are included in what is called dispersion corrections taken in ref. \cite{Ericson:2000md} from ref. \cite{Mizutani:xw,Afnan:1974ye}. 

We consider double scattering with one $\pi N$ vertex in $s$--wave and the other one in $p$--wave and viceversa, or the two vertices in $p$--wave. After performing the boosts to the CM frames we find
\be
a_{\pi^-d}^{(s-p)}&=&4\pi\;\frac{\left(1+m_\pi/m_N\right)^2}{1+m\pi/m_d}\;\left(2b_0c_0-4b_1c_1\right)
\int\frac{d^3{\bf q}}{(2\pi)^3}\;\frac{1}{{\bf q}^2}\int d^3{\bf p}\;{\tilde \varphi}({\bf p})\;{\tilde \varphi}({\bf p+q})^\star \non
&&\left[\left(\frac{m_\pi}{m_\pi+m_N}\right)^2\left({\bf p}^2+({\bf p+q})^2\right)-\frac{m_\pi}{m_\pi+m_N}\;{\bf q}^2\right]
\label{fermi_sp}
\ee
with ${\tilde \varphi}({\bf p})$ the Fourier transform of the deuteron wave function $\varphi({\bf r})$, normalized such that $\int d^3{\bf p}\;|{\tilde \varphi}({\bf p})|^2=1$. The large term in eq. (\ref{fermi_sp}) comes from the ${\bf q}^2$ of the square bracket in the equation. This term is easily evaluated since it becomes proportional to $|\varphi({\bf r}=0)|^2$. The precise value of $\varphi(0)$ is not well known since it depends on the short range forces assumed in the model. Thus one can imagine that there will be large uncertainties in this term. For the $d$--wave part of the wave function it vanishes, but the $s$--wave part provides a contribution, as can be seen in Tab. \ref{tab:fermi_correction},
\linespread{1.3}
\begin{table}
\caption{Fermi corrections to $a_{\pi^- d}$ in double scattering.}
%\begin{tabular*}{0.7\textwidth}{@{\extracolsep{\fill}}|l|l|l|l|l|}
\begin{tabular}{|l|l|l|l|l|}
\hline
{\bf Contribution}&\multicolumn{2}{|c|}{Bonn from ref. \cite{Machleidt:2000ge}}&\multicolumn{2}{|c|}{Paris from ref. \cite{Lacombe:eg}}
\\ 
$[10^{-4}\cdot m_{\pi}^{-1}]$&$s$\hspace*{1.2cm}&$d$&$s$\hspace*{1.2cm}&$d$
\\ \hline
${\bf q}^2$--term of eq. (\ref{fermi_sp})&$-30$&$0$&$-2$&$0$\\
Rest of eq. (\ref{fermi_sp})&$13$&$9$&$11$&$13$\\
From eq. (\ref{fermi_pp})&$-3$&$2$&$-4$&$2$\\ 
\hline \hline
{\bf Sum}&\multicolumn{4}{|c|}{$5\pm 15$}
\\ \hline
\end{tabular}
\label{tab:fermi_correction}
\end{table}
\linespread{1.0}
which depends much on the model. It is interesting to notice that in the case of the Paris potential where there is a stronger repulsion at short distances, the value of the correction is much smaller than that for the Bonn potential. We do not consider here the range of the $s$--wave part. It does not come as in the impulse approximation but it produces a small correction to another small correction. 

As for the $p$--wave in both vertices the contribution is
\be
a_{\pi^-d}^{(p-p)}&=&4\pi\;\frac{\left(1+m_\pi/m_N\right)^2}{1+m\pi/m_d}\;\left(2c_0^2-4c_1^2\right)
\int\frac{d^3{\bf q}}{(2\pi)^3}\;\frac{1}{{\bf q}^2}\int d^3{\bf p}\;{\tilde \varphi}({\bf p})\;{\tilde \varphi}({\bf p+q})^\star \non
&&\left[\left(\frac{m_\pi}{m_\pi+m_N}\right)^2 ({\bf p+q})^2-\frac{m_\pi}{m_\pi+m_N}\;{\bf (p+q)q}\right] \left[\left(\frac{m_\pi}{m_\pi+m_N}\right)^2{\bf p}^2+\frac{m_\pi}{m_\pi+m_N}\;{\bf p q}\right].
\label{fermi_pp}
\ee
This contribution should be smaller than the former one by comparing the strength of double scattering in $s$--wave with double scattering with $s$ and $p$--waves. A straightforward evaluation with a monopole form factor in each vertex with $\Lambda=1$ GeV, since beyond that momentum, the wave function is certainly unreliable, gives indeed a small contribution compared to the typical size of the corrections discussed. The results obtained with eq. (\ref{fermi_pp}) for the double scattering with $p$--wave, shown in Tab.  \ref{tab:fermi_correction}, agree with the value of $-3$ in the same units quoted in refs. \cite{Baru:xf,Ericson:2000md}. 

\vspace*{0.3cm}

{\it Isospin violation}

\vspace*{0.1cm}

The topic is thoroughly investigated in \cite{Fettes:1998wf} but not considered in \cite{Beane:2002wk} for the evaluation of the $\pi N$ scattering lengths from pionic atom data. Their non consideration reverts into admittedly smaller errors in $(b_0, b_1)$ than the given ones according to ref. \cite{Beane:2002wk}. The effects from this source are estimated relatively small in the $\pi^- d$ scattering length, of the order of $3.5\cdot 10^{-4} m_\pi^{-1}$ according to \cite{Ericson:2000md,Baru:xf,Tarasov:yi}. We shall take a different attitude. Our approach allows for isospin violation since the masses of the particles are taken different. This is not the only source of isospin violation \cite{Fettes:1998ud,Fettes:1998wf,Gibbs:1995dm,Gibbs:1997jv,Matsinos:1997pb,Piekarewicz:1995tx}, but it gives the right order of magnitude. 
In order to consider breaking from other sources than mass splitting, we shall allow the subtraction constants $\alpha_i$ in the three $\pi N$ channels to be different. The global fit from the section \ref{sec_results} takes into account the isospin breaking in the $\pi^- d$ scattering length.

\vspace*{0.3cm}

{\it Non--localities of the $\pi N$ $s$--wave interaction}

\vspace*{0.1cm}

These are corrections to the assumption of point--like interaction in the $\pi N$ vertices. They are considered for single and double scattering in ref. \cite{Ericson:2000md} leading to a modification of the pion propagator $G\sim 1/r$. The non--locality of $\pi N$ interaction affects mainly the isovector part of $s$--wave $\pi N$ scattering \cite{Ericson:2000md}. This is closely associated with the VMD assumption, which states that the $\pi N$ interaction is predominantly mediated by the $\rho$---meson. In the work of ref. \cite{Inoue:2001ip} the $\rho$ meson is explicitly taken into account, modifying the Weinberg--Tomozawa term at intermediate energies, but close to pion threshold the $\rho$ does not play a role in the approach of \cite{Inoue:2001ip} and in the present work. Therefore, we adopt the corrections from refs. \cite{Ericson:2000md,Baru:xf,Tarasov:yi} where values of $17(9)\cdot 10^{-4} m_\pi^{-1}$ and $29(7)\cdot 10^{-4} m_\pi^{-1}$ are obtained, respectively. We shall take a value of $23\cdot 10^{-4} m_\pi^{-1}$, and a larger uncertainty of 15 in the same units, in order to account for the discrepancy of these two results.

\vspace*{0.3cm}

{\it Nonstatic effects}

\vspace*{0.1cm}

These are corrections that go beyond the assumptions made from fixed centers. The accuracy of the static approximation is further supported by the study of \cite{Baru:2004kw} where, due to the dominance of the isovector $\pi N$ amplitudes, recoil corrections are shown to be small. Estimates of these nonstatic effects are done in \cite{Faldt:1974sm,Baru:xf,Tarasov:yi} and, as quoted in \cite{Ericson:2000md}, they lead to a correction of $11(6)\cdot 10^{-4} m_\pi^{-1}$. Since the 5th diagram in Fig. \ref{Crossed_Diagram} is part of the nonstatic effects, we do not include its contribution in the corrections, but adopt the value from \cite{Faldt:1974sm,Baru:xf,Tarasov:yi}.

\vspace*{0.3cm}

{\it Dispersion corrections and other real parts of the amplitude}

\vspace*{0.1cm}

The dispersion corrections tied to the absorption of Figs. \ref{charge_states_in_absorption} and \ref{charge_man} have been calculated, e. g., in refs. \cite{Afnan:1974ye,Mizutani:xw,Thomas:1979xu,Tarasov:yi}. In ref. \cite{Afnan:1974ye}, a repulsive contribution of $-50\pm 3\cdot 10^{-4} m_\pi^{-1}$ is found (the error taken from the precision of displayed decimal digits in ref. \cite{Afnan:1974ye}). The authors of ref. \cite{Mizutani:xw} consider any absorption contribution to the real part from reactions of the type $\pi^- d\to NN\to \pi^- d$ within their non--relativistic treatment of the pion. In the formulation of the present study, this would correspond to the sum of the absorption diagrams of Figs. \ref{charge_states_in_absorption} and \ref{charge_man}, plus diagrams that contain external pions which couple directly via $p$--wave to a nucleon. The latter contribute only when Fermi motion is considered (see above). Their diagram $D$ would correspond to the diagrams of Fig. \ref{charge_states_in_absorption} in the present work, and $D'$ would correspond to the diagrams in Fig. \ref{charge_man}. Using different models and wave functions, the authors of ref. \cite{Mizutani:xw} obtain values for the sum of the diagrams $D$ and $D'$ of $+14.9$, $-19.5$, and $-9.28$ in the units of $10^{-4} m_\pi^{-1}$. This tells us that there are large intrinsic uncertainties in this calculation.

In ref. \cite{Thomas:1979xu}, a value for $\Delta a_{\pi^- d}$ of $-56\pm 14\cdot 10^{-4} m_\pi^{-1}$ was deduced from the two  publications \cite{Afnan:1974ye,Mizutani:xw}. The value of ref. \cite{Thomas:1979xu} is then adopted by the recent work of ref. \cite{Ericson:2000md}. The other important diagram included in ref. \cite{Mizutani:xw}, called $C$, which leads to this final number,  can be interpreted in a Feynman diagrammatic way as double scattering involving one $\pi N$ vertex and a second scattering involving only the $p$--wave mediated by the nucleon pole. According to the work done here this should be complemented also with the $\Delta$ pole, which is dominant, and the crossed nucleon pole term. Altogether, this would be the Fermi correction to double scattering with $s$ and $p$--waves evaluated before, and in some case (see Tab. \ref{tab:fermi_correction}) we found a correction of $-30\cdot 10^{-4} m_\pi^{-1}$, which follows the trend of the result of ref. \cite{Thomas:1979xu}, but we also discussed that this contribution is very uncertain since it depends on the unknown value of $\varphi(0)$ which is highly model dependent. 

We, hence, follow our Feynman diagrammatic technique accounting for the mechanisms implicit in refs. \cite{Afnan:1974ye,Mizutani:xw} and substitute their numbers by our dispersion correction plus the rescattering terms involving $p$--waves discussed above in the subsection of Fermi corrections.

Other possible contributions related to pion interaction with the pion cloud are shown to cancel with related vertex contributions in ref. \cite{Beane:2002wk}, something also found in $K^+$ and $\pi$ nucleus scattering in refs. \cite{Meissner:1995vp} and \cite{Oset:nr}, respectively.

\vspace*{0.3cm}

In addition, there are other minor contributions from the literature to the real part of the pion deuteron scattering length, and we show them in Table \ref{summary_literature} without further comments.

\linespread{1.3}
\begin{table}
\caption{Corrections to Re $a_{\pi^- d}$.}
\begin{tabular*}{0.7\textwidth}{@{\extracolsep{\fill}}|l|l|l|}
%\begin{tabular}{|l|l|l|}
\hline
{\bf Contribution}&{\bf Value} in $10^{-4}\cdot m_{\pi^-}^{-1}$&{\bf Source}
\\ \hline
$(\pi^- p,\gamma n)$ double scattering&$-2$&\cite{Ericson:2000md}\\
Form factor/ Non--locality&$23\pm 15$&\cite{Ericson:2000md,Baru:xf,Tarasov:yi}\\
Non--static&$11\pm 6$&\cite{Ericson:2000md}(\cite{Faldt:1974sm,Baru:xf,Tarasov:yi})\\
virtual pion scattering&$-7.1\pm 1.4$&\cite{Beane:1997yg}\\
Dispersion&$2.4\pm 4.3$&Present study\\
Crossed $\pi$ and $\Delta(1232)$&14.6&Present study\\
Fermi motion, IA&$57\pm 9$&Present study\\
Fermi m. double scatt.  (s-p,p-p)&$5\pm 15$&Present study
\\ \hline \hline
{\bf Sum}&$104\pm 24$&
\\ \hline
\end{tabular*}
\label{summary_literature}
\end{table}
\linespread{1.0}

\newpage
\section{Results}
\label{sec_results}
The threshold data from pionic hydrogen and deuterium are obtained from the PSI experiments of \cite{Schroder:uq,Schroder:rc,Sigg:qd,Sigg:1995wb}. According to preliminary results from PSI experiments \cite{Gasser:2004ub}, these results could change in the near future with a consequence in the extracted values of $b_0$, $b_1$. Systematic uncertainties were also considered in \cite{Ericson:2000md}. In addition, the Coulomb corrections on the pionic hydrogen have been recently revised in ref. \cite{Ericson:2003ju}. This shifts the hydrogen data by the order of one $\sigma$. We take the newer values from ref. \cite{Ericson:2003ju} but keep the more conservative error estimates from refs. \cite{Sigg:qd,Schroder:rc,Hauser:yd,Ericson:2000md}.
\be
a_{\pi^-p\to\pi^-p}&=&\left(870\pm 2\;\mbox{stat.}\;\pm 10\;\mbox{syst.}\;\right)\cdot 10^{-4} m_\pi^{-1}\non
a_{\pi^-p\to\pi^0n}&=&-1250(60)\cdot 10^{-4} m_\pi^{-1}\non
a_{\pi^- d}&=&\left(-252\pm 5\;\mbox{stat.}\;\pm 5\;\mbox{syst.}\;+i\;63(7)\right)\cdot 10^{-4} m_\pi^{-1}
\label{exp_overview}
\ee
The sum of the corrections to the real part of the pion--deuteron scattering length from the sections \ref{sec_absorption_dispersion}, \ref{section_further}, and \ref{corrections} is
{\boldmath
\be
\left( 104\;\pm \;24\right)\cdot 10^{-4} m_\pi^{-1}.
\label{sum_corrections}
\ee }
The corrections in (\ref{sum_corrections}) are positive, which means that this additional attraction of the pion must be compensated for by a larger contribution from the multiple scattering series in order to give the experimental value in (\ref{exp_overview}).

We proceed now to determine the isoscalar and isovector scattering length $(b_0,b_1)$ from the experimental data in (\ref{exp_overview}), and from low energy $\pi N$ scattering data \cite{Arndt:2003if}.
As Table \ref{table_one} shows, the deuteron scattering length is particularly sensitive to $b_0$. This is due to the impulse approximation that cancels in the limit $b_0\to 0$, but contributes significantly for $b_0\neq 0$ as the results for the phenomenological Hamiltonian (\ref{pheno_la}) in Table \ref{table_one} demonstrate. The data from pionic hydrogen, $a_{\pi^-p\to\pi^-p}$ and $a_{\pi^-p\to\pi^0n}$, on the other hand, provide exact restrictions on the isovector scattering length $b_1$. Finally, the low energy $\pi N$ scattering data, together with the threshold values, determine the free parameters of the coupled channel approach of section \ref{sec_unitary_model}.

The authors of refs. \cite{Lyubovitskij:2000kk,Gasser:2002am} use a different approach, based on effective field theory (EFT) in the framework of QCD and QED, which gives somewhat different values for the $\pi^-p$ scattering length, with larger errors tied basically to the poorly known $f_1$ parameter. That means, for instance, that these authors explicitly take into account corrections that go into the generation of mass splittings, while the empirical analyses that lead to the data in eq. (\ref{exp_overview}) and scattering data at finite energies, that we shall use later on, only deal with electromagnetic corrections through the Coulomb potential and use physical masses. Hence, the simultaneous use of data extracted through these different methods for the global analysis that we use here, should be avoided. A reanalysis of raw $\pi N$ and $\pi d$ data through the EFT techniques is possible, and steps in this direction are already given in \cite{Lyubovitskij:2000kk,Gasser:2002am,Fettes:2001cr}. In any case, we shall show later on how our final results change with different values of this $\pi^- p$ scattering length. The global analysis of the bulk of threshold and scattering data that we do in the present work leads us to use the phenomenological multiple scattering method which has been used to extract the amplitudes, which implies that the results we obtain should only be used within this framework.

\subsection{The isoscalar and isovector scattering lengths}
 
\subsubsection{Previous Results}
 
In the literature, different values for $(b_0, b_1)$ have been extracted from pionic atoms and low energy $\pi N$ scattering extrapolated to threshold. In ref. \cite{Koch:bn}, a value of $a_+=-80\pm 20\cdot 10^{-4} m_\pi^{-1}$ has been extracted from the data by performing a partial--wave analysis. Note, however, the comments in ref. \cite{Sigg:qd} on this value, concerning the outdated database and other uncertainties. 

Extrapolations of low energy $\pi N$ scattering to threshold have been updated over the years \cite{Arndt:bu,Arndt:1995bj,Arndt:2003if}, and the value for the isospin even scattering length in the SM95 partial wave analysis is $a_+=-30\cdot 10^{-4} m_\pi^{-1}$. In an earlier publication in ref. \cite{Arndt:bu}, the same group found a deep minimum in their global fit for $a_+=-100\cdot 10^{-4} m_\pi^{-1}$. The current value is $a_+=-10(12)\cdot 10^{-4} m_\pi^{-1}$ in the most recent analysis, FA02, of ref. \cite{Arndt:2003if}.

From the constraints of the strong shifts in hydrogen and deuterium pionic atoms, the authors of refs. \cite{Sigg:qd, Sigg:1995wb} deduce small and positive values for $a_+$ of $0$ to $50\cdot 10^{-4} m_\pi^{-1}$. The findings in ref. \cite{Sigg:qd} are still compatible with isospin symmetry, although the two bands (constraints) from the shift at a $\chi^2$ of 1 do barely intersect with the constraint from the hydrogen width. As pointed out in ref. \cite{Sigg:qd}, this would be evidence of isospin violation. The value in ref. \cite{Sigg:qd} of ($b_0,b_1$) relies on the corrections to the real part of $a_{\pi^- d}$ from the analysis of ref. \cite{Thomas:1979xu}. In a recent publication on new measurements of $a_{\pi^- d}$, ref. \cite{Schroder:rc}, one finds an extensive discussion on updated corrections, including refs. \cite{Ericson:2000md,Beane:1997yg,Baru:xf,Tarasov:yi}. The authors in \cite{Schroder:rc} find a value of $(b_0,b_1)=-1^{+9}_{-21}\cdot 10^{-4} m_\pi^{-1}$. See the discussion in ref. \cite{Ericson:2000md} on this value. The recent theoretical approach in ref. \cite{Deloff:2001zp} provides a value of around $b_0=(-30\pm 40)\cdot 10^{-4} m_\pi^{-1}$.

To conclude, the experiments and subsequent analyses on pionic atoms lead to a value of $b_0$ being compatible with zero, or slightly negative, with errors of the same size or much larger than $b_0$ itself. In the various extrapolations of low energy $\pi N$ scattering data to threshold, more negative values of $b_0$ are favoured.
 
\subsubsection{Constraints on $(b_0,b_1)$ from threshold data} 
 
The restrictions on $(b_0,b_1)$ can be separately analyzed for the three threshold data points of eq. (\ref{exp_overview}), and for low energy $\pi N$ scattering. The separation of threshold and finite energy allows for a consistency test of the data, and, on the other hand, for a test of the freedom that the theoretical model has, when only one threshold point is fitted.
The influence of the data in eq. (\ref{exp_overview}) corresponds to bands in the $(b_0,b_1)$ plane, whose width is determined by the experimental and theoretical errors. 

\begin{figure}
\begin{picture}(200,300)
\put(-110,0){
\includegraphics[width=13cm]{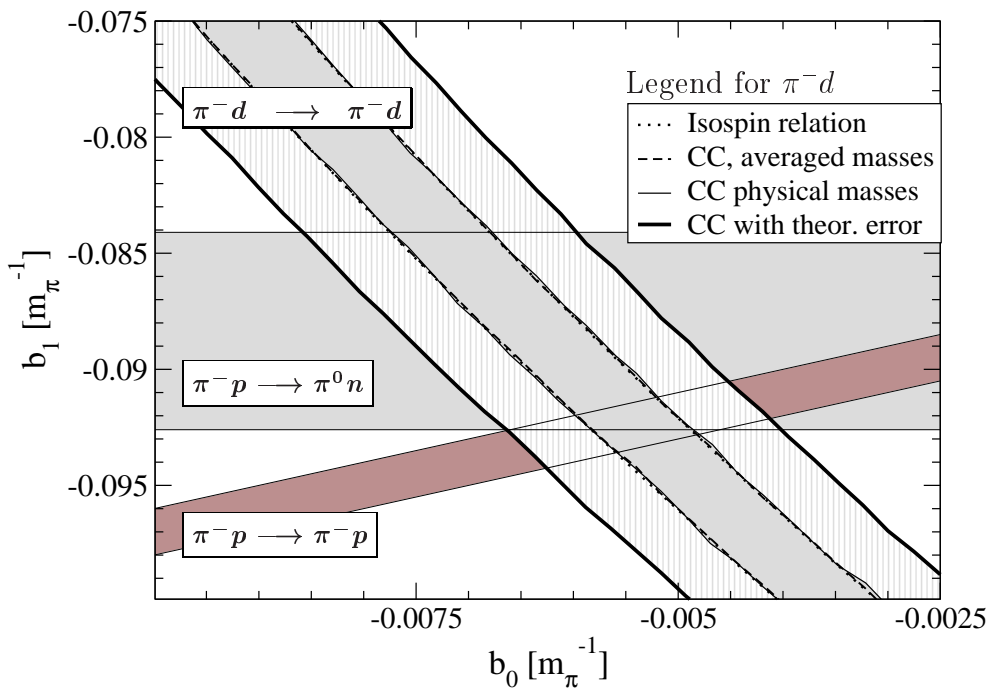}}
\end{picture}
\caption{(Color online) Constraints on $(b_0,b_1)$ from Re $a_{\pi^- d}$ and pionic hydrogen.}
\label{deuterium_band}
\end{figure}

We begin with the influence of $a_{\pi^- d}$ on $(b_0,b_1)$. For the real part of the deuterium scattering length $a_{\pi^- d\to\pi^- d}$ the band in the $(b_0,b_1)$ plane is calculated in four different ways, in order to determine the effect of isospin breaking by using physical masses instead of averaged ones. In all approaches, the large correction from eq. (\ref{sum_corrections}) is taken into account, and in the approaches one to three only the experimental error is considered. In the fourth approach, the large theoretical error from eq. (\ref{sum_corrections}) is added, widening the band by a factor of 3. In Fig. \ref{deuterium_band}, the results for the deuterium are plotted.
\begin{itemize}
\item
First, the four $\pi N$ scattering lengths in the solution of the Faddeev equations in eq. (\ref{analytic_kamalov}) are expressed in terms of $b_0$ and $b_1$. This, of course, implies the assumption of isospin symmetry. Then, random values of $(b_0,b_1)$ are generated and $a_{\pi^- d}$ is calculated with the help of eqns. (\ref{analytic_kamalov},\ref{folding_a}). The pairs $(b_0,b_1)$, that lead to a $\chi^2\le 1$ with the experimental value $a_{\pi^- d,\;{\exp.}}$ of eq. (\ref{exp_overview}), are kept, and plotted in Fig. \ref{deuterium_band} (dotted line). 
\item
In a second approach, we generate the four scattering lengths with the coupled channel (CC) approach of section \ref{sec_unitary_model}. For that, we take random values for the five free parameters of the theory $(\alpha,\beta,\gamma,c_i)$, and also use averaged masses for pions and nucleons, an assumption that we will drop in the third approach. Then, from the scattering lengths, $(b_0,b_1)$ are calculated with the help of eq. (\ref{first_eqn}). At the same time, $a_{\pi^- d}$ is calculated with the help of the Faddeev equations (\ref{analytic_kamalov},\ref{folding_a}). The same selection rule of $\chi^2\le 1$ as in the first approach sorts out the $(b_0,b_1)$ pairs that are plotted in Fig. \ref{deuterium_band} with the dashed line. The result coincides exactly with the first approach. This is indeed expected since we take the same subtraction constant $\alpha_{\pi N}$ for all $\pi N$ channels.
\item
In a third approach, we use physical masses instead of averaged ones, and proceed otherwise exactly as in approach 2. The result, plotted with the thin solid line in Fig. \ref{deuterium_band}, shows a nearly identical result compared to the first two approaches. This gives us a measure to which extend isospin breaking effects from different masses can affect the result. The maximal effect of isospin breaking from this source cannot exceed a small fraction of the experimental error.
\item
The fourth approach takes into account the large theoretical error from eq. (\ref{sum_corrections}) of $24 \cdot 10^{-4} m_\pi^{-1}$ and follows otherwise the third approach. The band is widened significantly (thick solid line). In the following calculations we use this approach.
\end{itemize}
 
The constraints from pionic hydrogen are also plotted in Fig. \ref{deuterium_band}. For each of the two bands from pionic hydrogen, marked as '$\pi^-p\to \pi^-p$' and '$\pi^-p\to\pi^0 n$' in Fig. \ref{deuterium_band}, we have followed the approaches 1 to 4 as for the deuteron. The four approaches give identical results for each band, as before. Therefore, in Fig. \ref{deuterium_band} one finds only one band from $a_{\pi^- p\to \pi^- p}$ and one from $a_{\pi^- p\to \pi^0 n}$.

The horizontal band shows the constraint from the experimental $a_{\pi^- p\to \pi^0 n}$ that is directly related to the hydrogen width. We have the isospin relation 
\be
b_1=1/\sqrt{2}\left(\;a_{\pi^-p\to\pi^0 n}\pm \Delta a_{\pi^-p\to \pi^0 n}\right)
\label{isorelation_width}
\ee 
where the experimental error $\Delta$ is relatively large, leading to a wide band. The hydrogen shift is closely related to $a_{\pi^-p\to \pi^-p}$, which leads to the constraint 
\be
b_1=b_0-a_{\pi^-p\to \pi^-p}\pm \Delta a_{\pi^-p\to \pi^-p}
\ee
in the isospin limit.  

Taking exclusively the data from pionic hydrogen, values of $b_0$ from $-70\cdot 10^{-4}m_\pi^{-1}$ up to positive numbers are allowed from Fig. \ref{deuterium_band}. Then, the band from pionic deuterium is added which appears with a steep slope and narrows significantly the region of allowed values of $b_0$. The range of $b_0$ is now determined by the position and width of the deuterium band, namely by eqns. (\ref{exp_overview}) and (\ref{sum_corrections}). This shows the necessity of having revised and extended the corrections of the $\pi^- d$ scattering length in the former sections. Indeed, if we would not have applied the corrections from eq. (\ref{sum_corrections}), the deuterium band would show up in Fig. \ref{deuterium_band} with the same slope, but shifted by around $+55\cdot 10^{-4}m_\pi^{-1}$ along the $b_0$ axis. This would lead to a value of $b_0$ being perfectly compatible with 0. However, the situation after applying the corrections leaves us with a $b_0\in\left[-70,-40\right]\cdot 10^{-4}m_\pi^{-1}$.

\subsubsection{Pion Nucleon Scattering at finite energies}
The unitarized coupled channel approach is applied in order to describe $\pi N$ scattering at finite energies. 
We fix the free parameters of the theory by fitting the model to the data above threshold. Then, the threshold prediction of the model is calculated. This is called 'Extrapolation' in the following. Comparing the predictions at threshold with the experimental data from pionic atoms, eq. (\ref{exp_overview}), the low energy behavior and the consistency of the model is tested.

Additionally, it is desirable to have an accurate parametrization of the $\pi N$ amplitude over some energy range, from threshold up to moderate energies. This is achieved by including the threshold data themselves in the fit, and this is referred to as 'Global fit' in the following.

Selection of experimental data: From the analyses of the CNS data base \cite{Arndt:2003if} for $\pi N$ scattering we choose the 'single--energy solutions' values, which are obtained by fitting narrow regions in the CM energy $\sqrt{s}$ separately. In contrast to the global fit given in ref. \cite{Arndt:2003if}, the single energy bins carry individual errors each. This helps to determine the statistical influence of $\pi N$ scattering on the parameters of the model. We add a small constant theoretical error of 0.002 to the amplitudes in the normalization of \cite{Arndt:2003if}. The channels to be included are the $s$--wave isospin $I=1/2$ and $I=3/2$ amplitudes, with real and imaginary part. This does not mean four independent data points for each point in energy $\sqrt{s}$: Since the inelasticity is zero at the low energies included in all fits, real and imaginary part are totally determined by the phase shift $\delta$, and therefore, for the purpose of calculating the reduced $\chi_r^2$, there are only two independent values, from the $I=1/2$ and from the $I=3/2$ channel. 
\begin{figure}
\begin{picture}(200,300)
\put(-110,0){
\includegraphics[width=13cm]{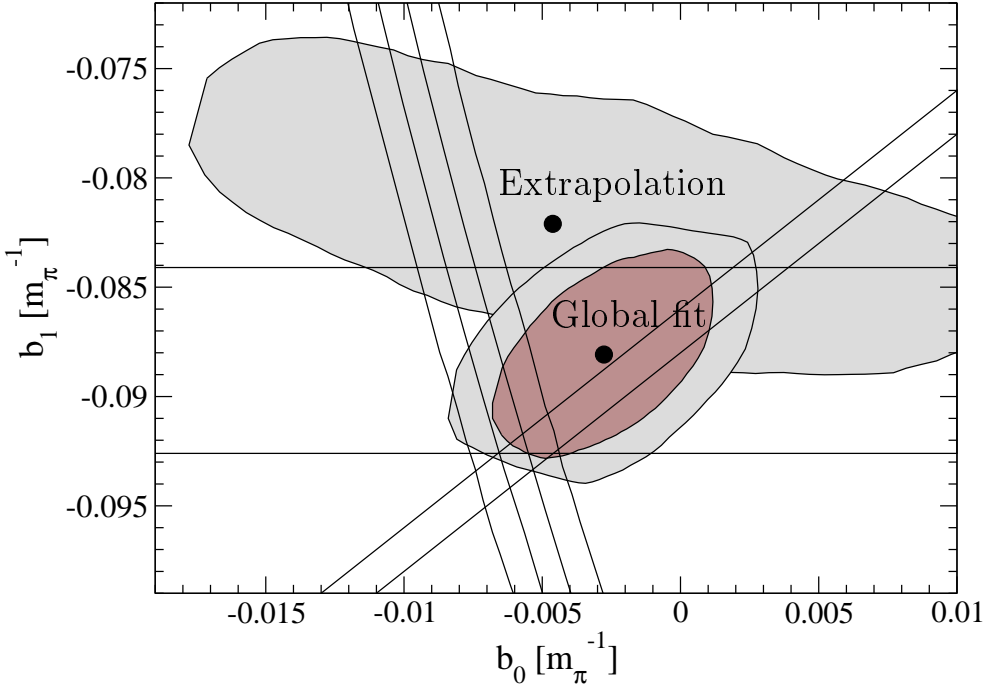}}
\end{picture}
\caption{(Color online) Threshold data from D and H, extrapolations from $\pi N$ scattering, and global fits.}
\label{all_constraints}
\end{figure}
In Fig. \ref{all_constraints}, the global fit and the extrapolation of $\pi N$ data to threshold are plotted. Table \ref{table_subtraction} displays the parameters of the fits, including the range of energies of the fitted data. 

\linespread{1.3}
\begin{table}
\caption{Global fits and threshold extrapolations from Fig. \ref{all_constraints}. Also, the fit without the damping factor from section \ref{sec_isoscalar} is displayed.}
\begin{tabular*}{0.9\textwidth}{@{\extracolsep{\fill}}|l||l|l||l|}
%\begin{tabular}{|l||l|l|l|l|}
\hline
&{\bf Global Fit}&{\bf Extrapolation}&{No damping}
\\ \hline
fitted data ($\sqrt{s}$)&1104--1253 MeV + threshold&1104--1253 MeV&1104--1180 MeV + threshold
\\ \hline
$\chi_r^2$&$51/(2\cdot 10+3)\simeq 2.2$&$24/(2\cdot 10)=1.2$&$33/(2\cdot 6+3)=2.2$
\\ \hline
$\alpha_{\pi N}$&$-1.143\pm 0.109$&$-0.990\pm 0.083$&$-1.528\pm 0.28$
\\
$2c_1-c_3$ [GeV$^{-1}$]&$-1.539\pm 0.20$&$-1.000\pm 0.463$&$-0.788 \pm 0.14$
\\
$c_2$ [GeV$^{-1}$]&$-2.657\pm 0.22$&$-2.245\pm 0.45$&$-1.670\pm 0.07$
\\
$\beta$ [MeV$^{-2}$]&$0.002741\pm 1.5\cdot 10^{-4}$&$0.002513\pm 3.3\cdot 10^{-4}$&No $\beta$
\\
$\gamma\; [10^{-5}\cdot m_\pi^{5}]$&$5.53\pm 7.7$&$-10\pm 6.1$&$10\pm 10$
\\ \hline
$\chi^2(a_{\pi^- p\to \pi^-p})$&$3$&$[91]$&$4$
\\
$\chi^2(a_{\pi^- p\to \pi^0 n})$&$<1$&$[2]$&$<1$
\\
$\chi^2(a_{\pi^- d})$&$8$&$[6]$&$4$
\\ \hline
\end{tabular*}
\label{table_subtraction}
\end{table}
\linespread{1.0}

The global fit favors values for $b_0$ and $b_1$ still negative but with smaller strength than the threshold data. As this fit contains threshold and finite energy data, it is situated between the extrapolation and the intersection of the three bands in the $(b_0,b_1)$ plane.
These differences between threshold and extrapolation from scattering data might be related to possible additional uncertainties in the phenomenological extraction of the partial wave amplitudes \footnote{G. H\"ohler, private communication.}.

In Fig. \ref{all_constraints}, the extrapolation and the global fit are indicated by shaded regions. These regions can be understood in the way that the reduced $\chi_r^2$ from Table \ref{table_subtraction} does not raise by more than 1 from the optimum ($\Delta\chi_r^2\le 1$) for all points $(b_0,b_1)$ inside these regions. For both fits, the optima are indicated with the black dots in Fig. \ref{all_constraints}. Furthermore, we plot for the global fit the region that fulfills $\Delta\chi_r^2\le 2$. It appears as the light grey area just around the $\Delta\chi_r^2\le 1$ region. 

The above explanation for the shaded regions of Fig. \ref{all_constraints} has to be taken with caution: The values $(b_0,b_1)$ inside the shapes have been {\it calculated} by the use of eq. (\ref{first_eqn}) from the elementary scattering lengths in the particle base. They are {\it not} the free parameters of the theory, which are the $\alpha$, $\beta$, $\gamma$, $c_2$, and $(2c_1-c_3)$. 

This implies that the $\Delta\chi_r^2\le 1,2$ criterion is applied to a $\chi^2$ that is a function of $a_{\pi^-p}$, $a_{\pi^-n}$, $a_{\pi^0 n}$, and $a_{\pi^- p\to \pi^0 n}$. Once a set of elementary scattering lengths $a_{\pi N}$ fulfills the criterion, $(b_0,b_1)$ are calculated from these values via eq. (\ref{first_eqn}), and give a point in the shaded regions of Fig. \ref{all_constraints}.

The elementary scattering lengths $a_i$ themselves have been calculated with the help of the CC approach by generating the fitting parameters  randomly in a wide range. Every set of $a_{\pi N}$ corresponds to a set $(\alpha,\beta,\gamma,c_i)$ and, via the criterion, the parameter errors on $(\alpha,\beta,\gamma,c_i)$ in Tab. \ref{table_subtraction} are determined.  'Parameter error' means here: The range of a parameter of a model, that leads to a raise of the reduced $\chi_r^2$ of less than 1 from the best $\chi^2$, minimizing $\chi^2$ at the same time with respect to all other parameters. 

As the final results for $(b_0,b_1)$ we take the values from the global fit:
{\boldmath
\be
\left(b_0,b_1\right)=\left(-28\pm 40,-881\pm 48\right)\cdot[10^{-4} m_\pi^{-1}].
\label{final_result}
\ee}
The errors have been taken from the maximal extension of the region in $(b_0,b_1)$, calculated from the $\Delta\chi_r^2\le 1$ criterion described above. They can be read off Fig. \ref{all_constraints}. The errors on $(b_0,b_1)$ take also into account the uncertainties from $\pi N$ scattering data up to $1253$ MeV.

With the caveat expressed in section \ref{sec_results} about using simultaneously data obtained through the EFT or phenomenological multiple scattering methods, we would also like to give, only as indicative, the results that we would obtain if we replaced the scattering length of $\pi^- p$ from eq. (\ref{exp_overview}) by the one given in ref. \cite{Gasser:2002am}. We find then the values
\be
\left(b_0,b_1\right)=\left(-39\pm 52,-862\pm 68\right)\cdot[10^{-4} m_\pi^{-1}].
\label{gasser_result}
\ee
which are compatible with the results in (\ref{final_result}) within the error bars.
The changes from eqns. (\ref{final_result}) to (\ref{gasser_result}) go in the same direction as in ref. \cite{Beane:2002wk} for $b_0$ but not for $b_1$. This is due to the fact that our fit puts more weight in the scattering data than that of ref. \cite{Beane:2002wk}.

\subsubsection{Finite energy behaviour of the fits}
\begin{figure}
\begin{picture}(200,360)
\put(-160,0){
\includegraphics[width=17cm]{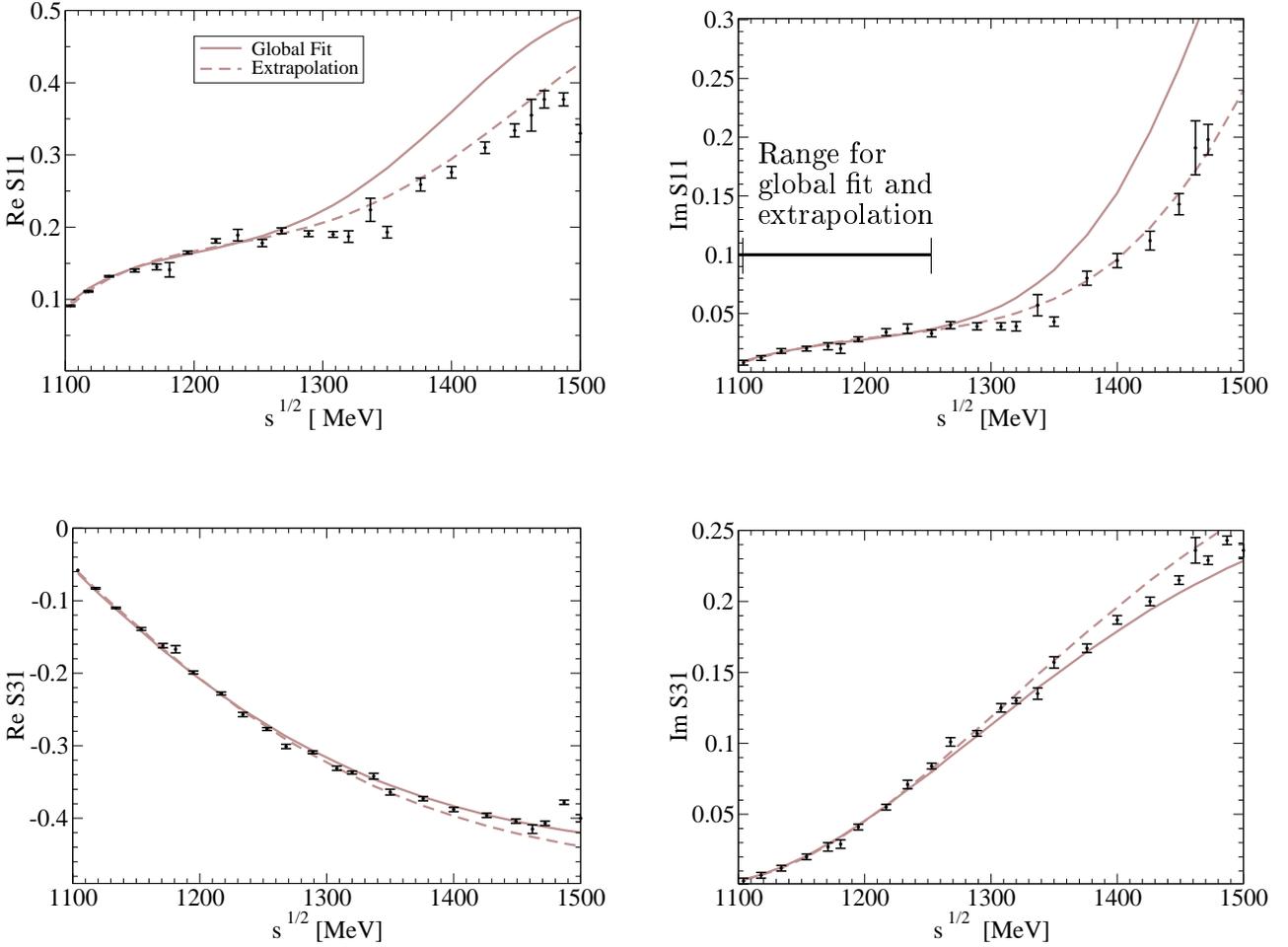}}
\end{picture}
\caption{(Color online) Global fits and extrapolations in the real and imaginary part of the  $I=1/2$ and $I=3/2$ channels. In the plot for \mbox{Im S11}, the range of fitted data for extrapolations and global fits is indicated.}
\label{fig_more}
\end{figure}

In Fig. \ref{fig_more} the energy behaviour of the global fit and the extrapolation from Fig. \ref{all_constraints} and Tab. \ref{table_subtraction} is displayed. The two upper pictures show the real and imaginary parts of the isospin $I=1/2$ channel, the lower the same for $I=3/2$. The data from the CNS in the single--energy solutions \cite{Arndt:1995bj} is displayed with errors. The global fit is displayed with the solid line, the extrapolation with the dashed line. 
As expected, the extrapolation provides better high energy behaviour, as it is not restricted by threshold data. It is remarkable, how well the extrapolation matches the data above 1253 MeV, which is the upper limit of the fitted data. 

In ref. \cite{Inoue:2001ip} the $\pi N$ scattering data has been fitted up to high energies, including the region of the $N^\star$ (1535) resonance. A fit was obtained that explained well the resonance but overestimated the $I=1/2$ amplitude at low energies, even when including the $\rho$--meson in the $t$--channel (see sec. \ref{sec_unitary_model}). The $\pi\pi N$ channel does not substantially improve the situation in ref. \cite{Inoue:2001ip}. It seems to be impossible to have at the same time a precise low energy fit, and a reproduction of the $N^\star(1535)$ resonance with the input of ref. \cite{Inoue:2001ip}. In the present approach we have introduced the extra isoscalar term which improves the fit of the low energy data. On the other hand, the present approach cannot reproduce the $N^\star$ (1535) resonance since the heavier members of the baryon and meson octet become important at these energies and one has to use the full $SU(3)$ approach as in ref. \cite{Inoue:2001ip}.

\subsubsection{The size of the isoscalar piece and the $(\pi,2\pi)$ term}
\label{sec_isoscalar}
\linespread{1.3}
\begin{table}
\caption{Isoscalar quantities: $\beta$, isoscalar generated by rescattering in $\pi N$, and final result of the fits.}
\begin{tabular*}{0.5\textwidth}{@{\extracolsep{\fill}}|l|l|l|}
%\begin{tabular}{|l||l|l|l|l|}
\hline
&{\bf Global fit}&{\bf Extrapolation}
\\ \hline
$b_c \;\;[10^{-4} m_\pi^{-1}]$&$-336$&$-434$
\\
$b_0\;\;[10^{-4} m_\pi^{-1}]$, generated&$442$&$396$
\\
$b_0\;\;[10^{-4} m_\pi^{-1}]$, final&$-28$&$-46$
\\ \hline
\end{tabular*}
\label{table_isos}
\end{table}
\linespread{1.0}
In table \ref{table_isos} we compare the two sources of isoscalar strength. In the first line we show the value of the contribution to $b_0$ from the isoscalar term of eq. (\ref{iso_lagrangian}), 
\be
b_c=-\frac{1}{4\pi}\;\frac{m_N}{m_\pi+m_N}\;\frac{4c_1-2c_2-2c_3}{f_\pi^2}\;m_\pi^2.
\ee 

The second line of Tab. \ref{table_isos} shows the $b_0$ that is generated by the rescattering of the $\pi N$ system. It has been calculated for the fits by setting all parameters except the subtraction constant $\alpha$ to zero. In this way, one can extract the size of the isoscalar part that is generated by the multiple loop sum from the Bethe--Salpeter equation (\ref{bs}). Although the lowest order chiral Lagrangian from ref. \cite{Inoue:2001ip} provides pure isovector interaction, the rescattering generates an isoscalar part, where usually 90 \% is generated by one loop, and most of the rest by the 2--loop rescattering (depending on the actual values of the subtraction constants). 
The last line of Tab. \ref{table_isos} provides the final value of $b_0$. The interplay of the isoscalar piece from eq. (\ref{iso_lagrangian}), the isovector interaction, and the subtraction constant leads to a resulting $b_0$ (last line of Tab. \ref{table_isos}) that cannot be explained any more as the sum of $b_c$ plus the generated $b_0$.

It is instructive to compare the results that we obtain for the isoscalar coefficients $c_i$ in Tab. \ref{table_subtraction} and those obtained in ref. \cite{Fettes:2000bb}. The results of the fit $2^\dagger$ of the Tab. 4 in ref. \cite{Fettes:2000bb} are:
\be
2 c_1-c_3&=&-1.63\pm 0.9\; {\rm GeV}^{-1}\non
c_2&=&-1.49\pm 0.67\; {\rm GeV}^{-1}.
\label{c1c2c3}
\ee
The agreement with the global fit from Tab. \ref{table_subtraction} is fair within errors. 

A more direct comparison with the results of ref. \cite{Fettes:2000bb} can be achieved by removing the damping factor $\beta$ in the isoscalar term. We can only get good agreement with data up to about $\sqrt{s}=1180$ MeV. Restricting ourselves to energies below $\sqrt{s}=1180$ MeV, the fit to the data provides the parameters given in Tab. \ref{table_subtraction} as 'No damping'. One can see that the $c_2$ coefficient is in good agreement with eq. (\ref{c1c2c3}) and also the combination of $2c_1-c_3$ within errors. With this fit to the restricted data we find
\be
\left(b_0,b_1\right)=\left(-37\pm 37,-887\pm 43\right)\cdot[10^{-4} m_\pi^{-1}].
\ee 

It is  worth stressing this agreement since the ''standard'' values of the $c_i$ coefficients used in chiral perturbative calculations, where the $\Delta$ is not taken into account explicitly, are much larger in size than the coefficients found here or in ref. \cite{Fettes:2000bb} and lead to the combinations of eq. (\ref{c1c2c3}) with opposite sign. Given the large amount of problems originated by the use of the ''standard'' (large) $c_i$ coefficients in chiral perturbative calculations, the models to fit $\pi N$ cross sections including explicitly the $\Delta$, as in ref. \cite{Fettes:2000bb}, and the ''smaller'' $c_i$ coefficients found here and in ref. \cite{Fettes:2000bb}, are highly recommendable. 

\vspace*{0.2cm}

As for the $\gamma$ parameter corresponding to the $\pi N\to\pi\pi N$ mechanism, we should expect on physical grounds quite a small contribution. Indeed, this is the case: The size of the $\pi N\to \pi N$ term including two $\pi N\to \pi\pi N$ vertices and the $\pi\pi N$ loop for the global fit 2 corresponds to about 5\% 
of the tree level $\pi N$ amplitude.

\subsubsection{Isospin breaking}
The isospin breaking in pion nucleon scattering has received much attention both phenomenologically and theoretically \cite{Gibbs:1995dm,Gibbs:1997jv,Matsinos:1997pb,Piekarewicz:1995tx,Fettes:1998ud,Fettes:1998wf}. Our theoretical model uses isospin symmetry up to breaking effects from the use of different masses. We also rely for the fit upon some scattering data that has been analyzed assuming isospin symmetry \cite{Arndt:2003if}. A possible measure of isospin breaking can be given by the quantity $D$ which describes the deviation from the triangle identity,
\be
D=f_{\rm CEX}-\frac{1}{\sqrt{2}}\;(f_+-f_-).
\label{eqd}
\ee
In ref. \cite{Gibbs:1995dm}, $D$ is calculated by fitting $s$ and $p$--wave of $f_+$, $f_-$, $f_{\rm CEX}$ which are the amplitudes of $\pi^+p$, $\pi^-p$, and $\pi^-p\to\pi^0n$. The authors fit a variety of potential models to the elastic channels, and predict $f_{\rm CEX}$ from that. Fitting in a second step $f_{\rm CEX}$ alone, they state a discrepancy between the prediction and the fit of  $f_{\rm CEX}$ that results in a value of $D=-0.012\pm 0.003\; {\rm fm}$. Physical masses are included in the coupled channel approach of ref. \cite{Gibbs:1995dm}, and their value for $D$ contains effects of other sources of isospin violation than mass splitting.

In the present work we have derived a microscopical model only for the $s$--wave interaction, also using physical masses in a coupled channels approach. Unfortunately, there is no partial wave analysis available that is free of isospin assumptions, so that we cannot evaluate $D$ from eq. (\ref{eqd}) using only experimental $s$--wave amplitudes. For this reason we shall take advantage of the work done in \cite{Gibbs:1995dm}. 

\linespread{1.3}
\begin{table}
\caption{Additional fits with $D$ and different $\alpha_{\pi N}$}
\begin{tabular*}{0.4\textwidth}{@{\extracolsep{\fill}}|l|l|}
%\begin{tabular}{|l||l|l|l|l|}
\hline
&D 3 $\alpha$
\\ \hline
$\chi_r^2$&$2.3$
\\
$D$ [fm] at 1110 MeV&$-0.011$
\\ \hline
$\alpha_{\pi^-p}$&$-0.717\pm 0.59$
\\
$\alpha_{\pi^0n}$&$-0.823\pm 0.26$
\\
$\alpha_{\pi^+p}$&$-1.130\pm 0.10$
\\
$2c_1-c_3$ [GeV$^{-1}$]&$-1.752\pm 0.22$
\\
$c_2$ [GeV$^{-1}$]&$-2.676\pm 0.27$
\\
$\beta$ [MeV$^{-2}$]&$0.002731\pm 1.8\cdot 10^{-4}$
\\
$\gamma\; [10^{-5}\cdot m_\pi^{5}]$&$10.0\pm 9.0$
\\ \hline
\end{tabular*}
\label{more_fits}
\end{table}
\linespread{1.0}

The global fit from Tab. \ref{table_subtraction}, which only contains an isospin violation from mass splitting, produces $D=-0.0066\;\;{\rm fm}$, around half the value of ref. \cite{Gibbs:1995dm}. 
In order to give more freedom to the model to violate isospin symmetry from different sources than mass splitting, we now allow different subtraction constants $\alpha_i$ in each channel, $\pi^-p$, $\pi^0n$, and $\pi^+ p$. Then, we add extra data points taking the value of $D$ from ref. \cite{Gibbs:1995dm} at three energies covering the range of $T_{\rm lab}=30-50$ MeV (as in Fig. 1 of ref. \cite{Gibbs:1995dm}). 
The $\chi^2$ stays practically the same compared to the global fit in Tab. \ref{table_subtraction}, but the fit gives $D=-0.011\;{\rm fm}$ at $\sqrt{s}=1110$ Mev, see 'D 3 $\alpha$' in Tab. \ref{more_fits}. The scattering amplitudes from this fit are practically the same as those shown in the Figs. \ref{all_constraints} and \ref{fig_more} for the global fit and we do not plot them again.

As we can see in Tab. \ref{more_fits} the subtraction constant for the $\pi^+ p$ channel barely changes with respect to the global fit in Tab. \ref{table_subtraction} while those for the $\pi^- p$ and $\pi^0 n$ channels are reduced in size by about 30 \%. The values that we obtain in this fit for $2c_1-c_3$ and $c_2$ change little with respect to those quoted before and the values for $(b_0,b_1)$ reveal a shift compared to the ones of the global fit in eq. (\ref{final_result}) of
\be
\delta b_0=6\cdot 10^{-4}\;m_\pi^{-1},\quad \delta b_1=9\cdot 10^{-4}\;m_\pi^{-1}.
\ee
The changes induced by the extra isospin breaking are rather small compared with the errors that we already have. 

The last value obtained for $D$ is in agreement with ref. \cite{Gibbs:1995dm}. A warning should be given about the solution found, since the isospin 1/2 amplitudes used in the fit imply isospin symmetry. However, the fact that half the amount of $D$ that we obtain comes from the use of different masses without invoking isospin breaking from other sources, and that the threshold data, which have small error bars, and thus a large weight in our fit, do not imply isospin symmetry, makes us confident that the solution obtained accounts reasonably for isospin violation in the problem. 
 
\section{Conclusions}
For a determination of the isoscalar and isovector scattering lengths of the $\pi N$ system, new calculations on the complex pion deuteron scattering length have been performed. The imaginary part of $a_{\pi^- d}$ shows a very good agreement with experimental data. The dispersive part from absorption has been found to be compatible with zero. This, together with corrections from crossed diagrams and the $\Delta(1232)$ resonance, and with other corrections taken from the literature, leads to a substantial shift of the real part of $a_{\pi^- d}$ towards positive values. 

The unitary coupled channel approach of ref. \cite{Inoue:2001ip} has been tested for consistency at low energies. However, we have added an isoscalar term that can be matched to terms of higher orders of the chiral Lagrangians. These terms are known to play an important role at threshold. With this additional ingredient to the model, together with the $\pi N\to 2\pi N$ channel, an acceptable global fit for the $\pi N$ amplitude up to intermediate energies has been obtained.

One of the results of the present work concerns the values of the $c_i$ parameters used in chiral perturbation theory at low energies. We find them compatible with values obtained from fits to data when the $\Delta$ is explicitly taken into consideration. On the other hand, we have addressed the isospin violation issue and found that our fit to data accounts for about half the isospin breaking only from mass splittings. The model has been extended to account for other sources of isospin breaking and then can match results of isospin breaking found in other works. We find that the effect of this breaking in the $b_0$, $b_1$ parameters is well within uncertainties from other sources. 

Attention to the sources of errors has been paid and we find larger values than in former studies. Altogether, we have here a new determination of the $\pi N$ scattering lengths and a new parametrization of the $\pi N$ $s$--wave amplitudes at low energies that can be used as input in studies of other elementary processes or as input to construct optical potentials from pionic atoms, where problems tied to the strength of the isoscalar part of the potential still remain.

\section*{Acknowledgments}
M. D. would like to thank the 'Studienstiftung des Deutschen Volkes' for financial support in the framework of his PhD thesis. This work is partly supported by DGICYT contract number BFM2003-00856,
and the E.U. EURIDICE network contract no. HPRN-CT-2002-00311. The authors would also like to thank T.E.O. Ericson for useful discussions and J. Gasser, G. H\"ohler, U. Mei\ss ner, and M.M. Pavan for valuable comments.

\appendix*
\section{The $d$--wave in the deuteron}
In the sections \ref{sec_absorption_dispersion} and \ref{section_further}, the influence of the $d$--wave and the interference of $d$--wave and $s$--wave have been discussed for a number of diagrams, namely the absorption, and the 5th diagram of Fig. \ref{Crossed_Diagram}. We display some explicit formulas for the coupling of spin and angular momentum involved in these calculations.

In momentum representation, the deuteron wave functions in eqns. (\ref{t_absorption}), (\ref{all_in_one}), and (\ref{eval_crossed}) are given by 
\be
F_d=F_d^{(s)}+F_d^{(d)},\quad F_d^{(i)}(p,\theta_p,\phi_p)=\left(2\pi\right)^{3/2}f_\nu^{(i)} ({\bf \hat{p}})\;\Psi^{(i)} (p), \quad \Psi^{(i)} (p)=\sqrt{\frac{2}{\pi}}\;\sum_{j=1}^{n}\;\frac{[C_j\;\mbox{for}\;i=s,\;D_j\;\mbox{for}\;i=d]}{p^2+m_j^2}.
\label{final_d}
\ee
The index $i$ stands for $s$ or $d$--wave, and the parametrizations of the radial part $\Psi (p)$, by means of the $C_j$, $D_j$, and $m_j$, are taken from ref. \cite{Machleidt:2000ge} $(n=11)$ for the CD--Bonn and ref. \cite{Lacombe:eg} $(n=13)$ for the Paris wave function. The normalization is $\int dp\;p^2\left(\Psi^{(s)}(p)^2+\Psi^{(d)}(p)^2\right)=1$ in both cases.  

The angular structure of the $d$--wave is given by the angular momentum $l=2,\;l_z=\mu$ and the spin wave function $\chi$
\be
f_\nu^{(d)}({\bf \hat{p}})=\sum_\mu C\;(2,1,1;\;\ \mu,\nu-\mu,\nu) Y_{2,\mu}(\theta_p,\phi_p)\;\chi^{S=1}_{S_z=\nu-\mu}, \quad f_\nu^{(s)}=Y_{00}\;\chi^{S=1}_{S_z=\nu}.
\label{ang_y}
\ee
The angular structure $f_\nu^{(i)}$, also normalized to one, is preserved under Fourier transforms from coordinate space. The index $\nu=-1,0,1$ indicates at which 3rd component $J_z$ the total angular momentum is fixed. The $C$'s in eq. (\ref{ang_y}) are the Clebsch--Gordan coefficients that couple the nucleon spins to the angular momentum $l=2$ of the $d$--state, in order to give a total angular momentum $J$ of one. The nucleons are necessarily in a spin triplet with $S=1$ for $s$ and $d$--wave.

Fixing in all calculations the third component of $J$ at $J_z=\nu=0$, we obtain for the $d-s$ interference of the diagrams of absorption in Fig. \ref{charge_states_in_absorption} 
\be
&&\left(\vec{\sigma}_{{\rm nucleon}\;1,2}\cdot{\bf q}\right) \left(\vec{\sigma}_{{\rm nucleon}\;1,2}\cdot{\bf q'}\right)\;f_0^{(d)}({\bf \widehat{{\bf q+l}}})f_0^{(s)}({\bf \widehat{{\bf q'+l}}})\non
&=&\frac{1}{2\sqrt{2}\pi}\;\left[1-3\cos^2 \theta_{\widehat{{\bf q+l}}}\right]{\bf q\cdot q'}\non
&+&\frac{3}{2\sqrt{2}\pi}\;\cos\theta_{\;\widehat{{\bf q+l}}}\sin\theta_{\;\widehat{{\bf q+l}}}\left[ \sin\phi_{\;\widehat{{\bf q+l}}}\left({\bf q'\times q}\right)_x-\cos\phi_{\;\widehat{{\bf q+l}}}\left({\bf q'\times q}\right)_y\right]
\label{first_int}
\ee
where symmetries in the amplitude $A$ in eq. (\ref{all_in_one}) have been used. The $\sigma$ matrices act on the same nucleon. The second line corresponds to the first term in the decomposition $\vec{\sigma}{\bf q'}\;\vec{\sigma}{\bf q}={\bf q'\cdot q}+i\left({\bf q}'\times{\bf q}\right)\cdot\vec{\sigma}$, and the third line to the term with crossed momenta. 

Another spin structure is given by the diagrams in Fig. \ref{charge_man} and the 5th diagram of Fig. \ref{Crossed_Diagram}. Here, the $\pi NN$ vertices are attached at different nucleons, and we obtain for the $d-s$ interference for Fig. \ref{charge_man}
\be
&&\left(\vec{\sigma}_{{\rm nucleon}\;1,2}\cdot{\bf q}\right) \left(\vec{\sigma}_{{\rm nucleon}\;2,1}\cdot{\bf q'}\right)\;f_0^{(d)}({\bf \widehat{{\bf q+l}}})f_0^{(s)}({\bf \widehat{{\bf q'+l}}})\non
&=&\frac{1}{2\sqrt{2}\pi}\;\left[1-3\cos^2 \theta_{\widehat{{\bf q+l}}}\right]\left(q_x q_x'+q_y q_y'-q_z q_z'\right)\non
&+&\frac{3}{2\sqrt{2}\pi}\;\cos\theta_{\;\widehat{{\bf q+l}}}\sin\theta_{\;\widehat{{\bf q+l}}}\left[ \sin\phi_{\;\widehat{{\bf q+l}}}\left(q_z' q_y+q_z q_y'\right)+\cos\phi_{\;\widehat{{\bf q+l}}}\left(q_z' q_x+q_z q_x'\right)\right].
\label{second_int}
\ee
The term $\left(q_x q_x'+q_y q_y'-q_z q_z'\right)$ in the second line of eq. (\ref{second_int}) is also present in the $s$-wave $\to$ $s$--wave transition and shows explicitly the $1/3$--contribution of the diagrams in Fig. \ref{charge_man} to the ones of Fig. \ref{charge_states_in_absorption}, as has been derived in a different way in eqns. (\ref{sigma_eins}), (\ref{sigma_zwei}), and (\ref{sigma_drei}).

For the 5th diagram of Fig. \ref{Crossed_Diagram}, one obtains the angular structure by replacing the momentum components $q_i\to q_i'$ in eq. (\ref{second_int}), where ${\bf q}'$ is the momentum of the disconnected pion (the $q$ in the angles $\theta_{\;\widehat{{\bf q+l}}}$ and $\phi_{\;\widehat{{\bf q+l}}}$ of eq. (\ref{second_int}) is not changed). 

The angular structure of $d$--wave $\to$ $d$--wave transition for the diagrams is calculated in analogy to eqns. (\ref{first_int}) and (\ref{second_int}), but the resulting expressions are lengthier due to the occurring double sums from eq. (\ref{ang_y}).

In the calculations for absorption, the results have been numerically tested for choices of $\nu$ equal to $\pm 1$ instead of 0, and they stay the same, as it is required.

\end{document}